\documentclass[aps,pre,twocolumn,groupedaddress]{revtex4}
\usepackage{latexsym}

\newenvironment{Fig}[1]{
\begin{figure}
\noindent\begin{minipage}[t]{\linewidth}
\begin{center}
\leavevmode
\epsfxsize=\linewidth
\epsfbox{#1}
}
{
\end{center}
\end{minipage}
\end{figure}
}

\begin{document}
\bibliographystyle{prsty}
\input epsf
\title{Structural Probe of a Glass Forming Liquid: Generalized
Compressibility}
\author{Herv\'e M. Carruzzo$^{\dagger}$ and Clare C. Yu}
\affiliation{
Department of Physics and Astronomy, University of California,
Irvine, Irvine, California 92697}
\date{\today}
\begin{abstract}
We introduce a new quantity to probe the glass
transition.  This quantity is a linear generalized compressibility which
depends solely on the positions of the particles.  We have performed a
molecular dynamics simulation on a glass forming liquid consisting of
a two component mixture of soft spheres in three dimensions. As the
temperature is lowered (or as the density is increased), the generalized 
compressibility drops 
sharply at the glass transition, with the drop becoming more and more
abrupt as the measurement time increases. At our longest measurement
times, the drop occurs approximately at the mode coupling temperature $T_C$.
The drop in the linear generalized compressibility occurs at the
same temperature as the peak in the specific heat. By examining the 
inherent structure energy as a function of temperature, we find that
our results are consistent with the kinetic view of the glass transition
in which the system falls out of equilibrium. 
We find no size dependence and no evidence for a second order phase transition
though this does not exclude the possibility of a phase transition below
the observed glass transition temperature.
We discuss the relation between the linear generalized compressibility
and the ordinary isothermal compressibility as well as the
static structure factor.
\end{abstract}

\pacs{PACS numbers: 64.70.Pf, 61.43.Fs, 02.70.Ns, 05.20.-y}
\maketitle

\section{Introduction}
The glass transition is still not well understood despite
extensive study. 
There have been two main theoretical approaches to the problem:
dynamic and thermodynamic. Theories in the first category
view the glass transition as a kinetic phenomenon
characterized by a growing relaxation time and viscosity
\cite{Ediger96,Fourkas97,Gotze92,Donati98,Yamamoto97b}.
When the relaxation time exceeds the measurement time, particle
motion appears to be arrested resulting in the glass transition.
One of the most prominent theories espousing this view
is the mode coupling theory (MCT) in which ideally the relaxation 
time diverges at a temperature $T_{C}$ above the experimental glass 
transition temperature \cite{Gotze92}.
Interesting and fruitful concepts such as 
dynamic inhomogeneities \cite{Donati98,Yamamoto97a,Tracht98} and the influence
of the energy landscape on relaxation processes \cite{Goldstein69,Sastry98}
have resulted from this approach.
The thermodynamic viewpoint attributes the glass transition to
an underlying phase transition hidden from direct
experimental observation by extremely long relaxation times
\cite{Ediger96,Fourkas97,Adam65,Gibbs58,Mezard99}.
In most scenarios there is an underlying second order phase
transition associated with a growing correlation length which produces
diverging relaxation times as well as diverging static susceptiblities
\cite{Kirkpatrick89,Sethna91,Kivelson95,Ernst91,Dasgupta91,Menon95}.
More recently Mezard and Parisi \cite{Mezard99,Parisi97b} have argued
that the underlying transition is actually a random first order
transition signaled by a jump discontinuity in the specific heat.

Experimentally the glass transition
is characterized by both kinetic and thermodynamic
features. For example in the supercooled liquid kinetic quantities
such as the viscosity and relaxation time grow rapidly as the
temperature is lowered. When the system falls out
of equilibrium below a certain temperature, thermodynamic
quantities exhibit features reflecting the 
glass transition. For example as the system is cooled the specific heat 
has a step--like form and the dielectric constant exhibits a peak
at a frequency dependent temperature.

In an effort to better characterize the glass transition we introduce
a novel probe which we call a generalized
compressibility \cite{Yvon58}. Unlike the specific heat which monitors energy
fluctuations, this linear compressibility is a function of the
microscopic structure of the system: it depends solely on the
positions of the particles and not on their previous history. 
Since we do not need to compare the system's state at different times,
it is not a dynamic or kinetic quantity. Rather 
it is a thermodynamic quantity in the sense that it is purely
a function of the microstate of the system dictated by its location 
in phase space. The generalized compressibility
is easy to compute numerically, and it is simpler than the
dielectric constant which involves both the translation and
orientation of electric dipoles. In addition it does not suffer from
finite size effects that can often plague measurements of the ordinary
compressibility deduced from simulations. The generalized compressibility
can be calculated in either 
the canonical or grand canonical ensembles. In particular it is well
defined for a system with
fixed volume $V$ and particle number $N$ in contrast to the ordinary 
compressibility which is defined for a system that has 
fluctuations in $N$ or $V$. The generalized compressibility
should be directly measurable experimentally in
colloidal suspensions of polystyrene spheres \cite{Weeks00} and
possibly in other systems as well.  In this paper we present 
measurements of this quantity in a molecular dynamics simulation of a
two component system of soft spheres. We find that the linear 
generalized compressibility drops sharply as the temperature
decreases below the glass transition temperature $T_g$. The drop
becomes more and more abrupt as the measurement time increases.
This is consistent with the structural arrest associated with a 
kinetic transition in which the system falls out of equilibrium.
Similar results are seen as the density is increased at fixed temperature.

The paper is organized as follows. Section II describes the molecular
dynamics simulations. Section III describes how the relaxation times
and mode coupling $T_C$ are determined. These are useful for setting
the time and temperature scales. Section IV describes our specific heat
measurements which show a peak at the glass transition.
Section V derives the expressions for
the linear and nonlinear generalized compressibilities, and shows our
results for these quantities. The linear generalized compressibility
shows an abrupt drop at the same temperature and density as the
peak in the specific heat. Section VI compares the ordinary isothermal
compressibility with our linear generalized compressibility and shows
the advantages of the latter. Section VII gives our results
for the diffusion constant. Section VIII explains the relation between
the linear generalized compressibility and the static structure factor.
Finally we summarize our results in Section IX.
A brief description of some of these results as well as results for
a single component fluid that forms a crystal was reported earlier
\cite{Carruzzo02}.

\section{Molecular Dynamics Simulation}
We have performed a molecular dynamics simulation on a
glass forming liquid \cite{Weber85,Kob94} consisting of a 50:50
binary mixture of soft spheres in three dimensions.
The two types of spheres, labelled A and B, differ only in their sizes.
The interaction between two particles a distance
$r$ apart is given by $V_{\alpha\beta}(r)
=\epsilon[(\sigma_{\alpha\beta }/r)^{12}+X_{\alpha\beta}(r)]$
where the interaction length
$\sigma_{\alpha\beta }=(\sigma_{\alpha}+\sigma_{\beta})/2$ with
$\sigma_B/\sigma_A=1.4$ ($\alpha$, $\beta =$ A, B).
For numerical efficiency, we set
the cutoff function $X_{\alpha\beta}(r)=r/\sigma_{\alpha\beta}-\lambda$
with $\lambda =13/12^{12/13}$. The interaction is cutoff at
the minimum of the potential $V_{\alpha\beta}(r)$.
Energy and length are measured in units of
$\epsilon$ and $\sigma_A$, respectively. Temperature is given in units of
$\epsilon /k_B$ where $k_B$ is Boltzmann's constant,
and time is in units of $\sigma_A\sqrt{m/\epsilon}$ where
$m$, the mass of the particles, is set to unity.
The equations of motion were integrated using the leapfrog
method \cite{Rapaport95} with a time step of 0.005.
During each run the average density $\rho_o=N/L^3$ was fixed,
and the temperature was kept constant using a constraint
algorithm \cite{Rapaport95}. The volume $V=L^3$.
$N=N_A+N_B$ is the total number of particles.
The system occupies a cube with dimensions
($\pm$ L/2, $\pm$ L/2, $\pm$ L/2) and periodic boundary conditions.

We have done sweeps of both temperature and density. We fix the parameters
so that crystallization is avoided upon cooling or when the density is
increased. For the temperature sweeps, we fix the density at $\rho_o=0.6$. 
For $\sigma_B/\sigma_A=1.4$, this corresponds to a packing fraction
of 1.04. Having a packing fraction larger than 1 means that
each particle was interacting with other particles most, if not all,
of the time. Our measuring procedure is the following.
For runs where we cool the system, we start each run at a high
temperature (T=1.5) and lower the temperature in steps of $\Delta
T=0.05$. At each temperature we equilibrate for $10^4$ molecular
dynamics steps (md steps) and then measure the quantities of interest
for $N_{\tau}$ additional md steps  where $N_{\tau}=10^5$, $2\times
10^5$, $10^6$, $3\times 10^6$, or $10^7$.  All the particles
move at each md
step. The results are then averaged over up to 40 different initial
conditions (different initial positions and velocities of the spheres).
We have done some runs in which we
heat the system of particles by starting at our lowest temperature
$T=0.1$ with a configuration obtained by cooling
the system. We then increased the temperature in steps of
$\Delta T=0.05$.  As before we equilibrate at each temperature for
$10^4$ time steps and then measure quantities for an additional $10^6$
time steps.

We have also done some density sweeps in which we fix the temperature
(T=1.0) and systematically change the density. The glass transition
occurs as we increase the density. Colloidal experiments
often study the glass transition as a function of density.
We start each run at a low density
($\rho=0.4$) and increase the density in steps of $\Delta\rho=0.025$.
At each density we equilibrate for $10^4$ md
steps and then measure the quantities of interest
for $N_{\tau}$ additional md steps.

The glass transition occurs either as the temperature is lowered or
as the density is raised. It is worth noting that temperature and 
density can be combined into a dimensionless parameter $\Gamma$
\cite{Barrat90}: 
\begin{equation}
\label{eq:gamma_eff}
\Gamma =\rho\sigma_{eff}^3/T^{\frac{1}{4}}\qquad
\sigma_{eff}^3=\sum_{\alpha\beta}n_{\alpha}n_{\beta}
\sigma_{\alpha\beta}^3
\end{equation}
where $\sigma_{eff}$ represents an effective diameter for particles in
the mixture. The concentration of each type
of particle is given by $n_A=N_A/N$ and $n_B=1-n_A$.
For our simulations $n_A=n_B=0.5$.
$\Gamma$ is the relevant parameter when
the particles spend most of their time sampling a nonzero interparticle
potential, i.e., for $\rho^{-1/3}<\sigma_{eff}$. Thus $\Gamma$ is
particularly useful for interparticle interactions which fall off 
with distance as a power law and do not have a cutoff beyond which
the interaction is zero. When cooling from the liquid phase, 
the glass transition is known to occur around $\Gamma=1.45$ \cite{Barrat90}.

We have looked for phase separation of the two types of spheres
by examining the distribution of large and small spheres in the 
neighborhood of large spheres and in the neighborhood of 
small spheres. We see no evidence for phase separation at either high 
($T=1.5$) or low ($T=0.15$) temperatures at a density of $\rho=0.6$.

\section{Relaxation Times and Mode Coupling $T_C$}
As points of reference for the time and temperature scales,
it is useful to find the mode coupling $T_C$ and the $\alpha$ relaxation
times. We can find the relaxation times using the intermediate
scattering function $F(\vec{k},t)$ which is a useful probe
of the structural relaxation. It is the spatial Fourier transform
of the van Hove correlation function $G(\vec{r},t)$ and the inverse
time transform of the dynamic structure factor $S(\vec{k},\omega)$.
There are two different types of intermediate scattering function: 
the self (incoherent) intermediate scattering function
$F_{s}(\vec{k},t)$ and the full (coherent) intermediate scattering function
$F(\vec{k},t)$.

In a computer simulation, the self (incoherent) part of the partial
intermediate scattering function $F_{s,\alpha}(\vec{k},t)$ can be 
calculated directly using
\cite{Hansen}
\begin{equation}
F_{s,\alpha}(\vec{k},t)=\frac{1}{N_{\alpha}}\left\langle
\sum_{i=1}^{N_{\alpha}}e^{i\vec{k}\cdot(\vec{r}_{i}(t)-\vec{r}_{i}(0))}
\right\rangle
\label{eq:SelfFkt}
\end{equation}
where the subscript $\alpha$ refers to the particle type, A or B.
$\vec{r}_{i}(t)$ is the position of particle $i$ at time $t$,
and $\langle ...\rangle$ refers to an average over different configurations.
The wave vector $\vec{k}=2\pi\vec{q}/L$ where $\vec{q}$ is a vector
of integers. For an isotropic system $F_{s,\alpha}(\vec{k},t)$ depends
only on the magnitude $k=|\vec{k}|$. We will choose $k=k_{max}$ where
$k_{max}$ is the position of the first maximum of the partial static
structure factor $S_{\alpha}(k)$. 
In Figure \ref{fig:selfFkt} we show the self intermediate scattering function
$F_{s,B}(k,t)$ versus time at temperatures below the caging temperature 
($T\approx 0.4$). The caging temperature is the highest temperature
at which a plateau is present in the intermediate scattering function
versus time.
The plateau represents the temporary localization of a particle by
a cage of other particles surrounding it.
\begin{Fig}{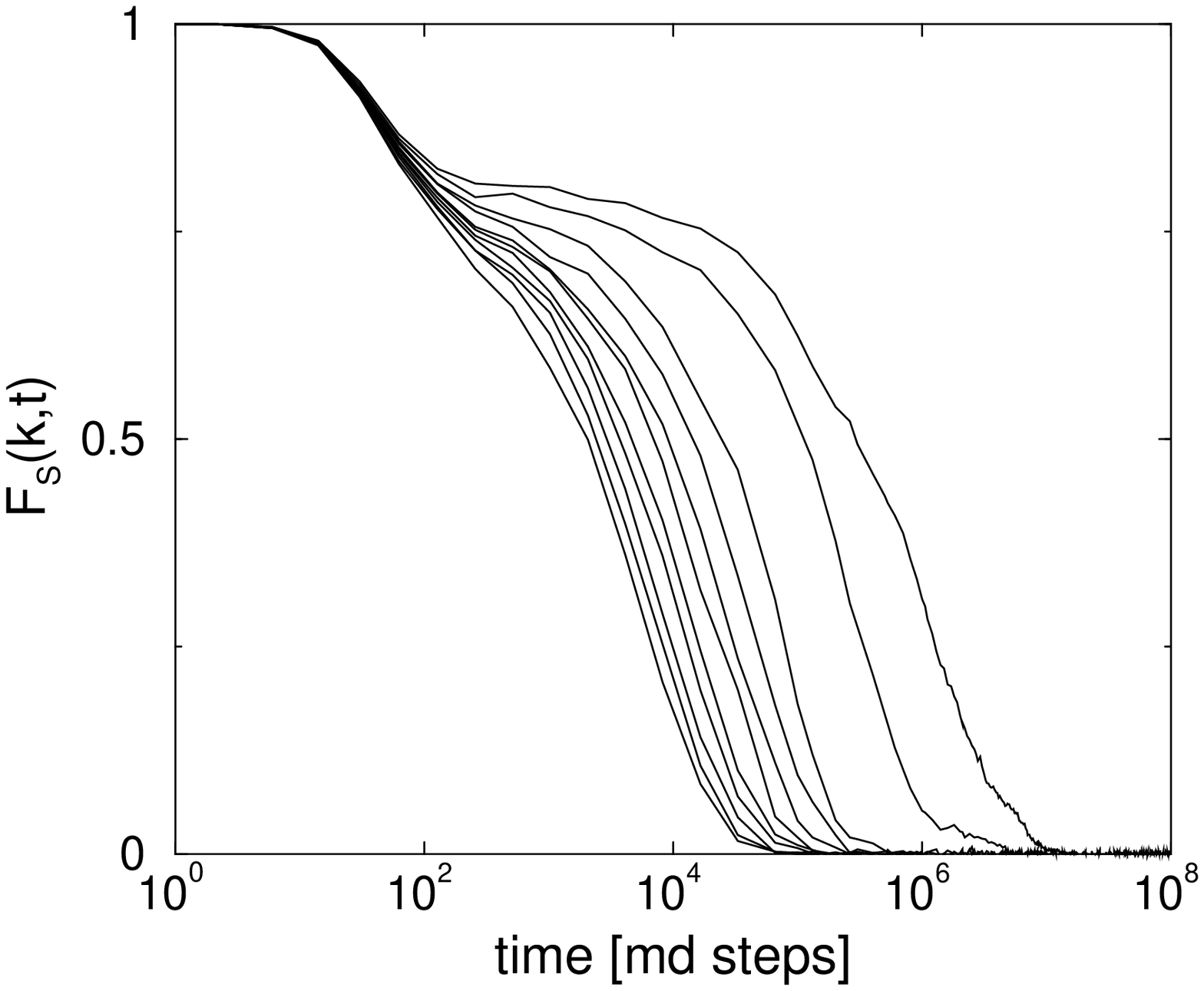}
\caption{The self intermediate scattering function versus time for a
system with 512 particles and $\rho_{o}=0.6$. The time is given in 
units of molecular dynamics time steps. From left to right, the curves are
for temperatures $T$ = 0.381679, 0.373134, 0.364964, 0.357143, 0.34965,
0.342466, 0.33557, 0.328947, 0.321543, 0.3021148, 0.289855 respectively. 
256 type B particles were used and the wave vector 
$k=2\pi \cdot 8.3666/L$ which is the location of the first peak in
the structure factor for type B particles. $L=8$. For each curve the system 
was initialized from
a configuration at that temperature obtained from parallel tempering
which is described in the Appendix.
Then the simulation was run only at that temperature. The
temperatures were chosen so that the parallel tempering acceptance rates were
high. The curves at the 7 highest temperatures were 
equilibrated for 1 million md time steps before recording the
configurations used to calculate $F_{s}(k,t)$. Each curve of the 7 highest
temperature curves is averaged over 24 runs except for $T=0.373134$ 
which is averaged over 54 runs. The curves for $T$ = 0.328947 and 0.321543
were equilibrated for 2 million md time steps before recording the
configurations used to calculate $F_{s}(k,t)$. These two curves were averaged
over 11 runs. The curve for $T$ = 0.3021148 was averaged over 22 runs and
was equilibrated for 10,000 md time steps before recording the configurations
used to calculate $F_{s}(k,t)$. The curve for $T$ = 0.289855 was averaged over
36 runs and equilibrated for 50 million md time steps before recording
configurations used to calculate $F_{s}(k,t)$.} 
\label{fig:selfFkt}
\end{Fig}

Mode coupling theory is applicable in the temperature range below
the caging temperature and somewhat above the mode coupling $T_C$.
We define the relaxation time $\tau_s$ by $F_{s}(k,\tau_s)=1/e$. We
determine the relaxation times for the seven highest
temperatures shown in Figure \ref{fig:selfFkt} and then 
fit the temperature dependence of $\tau_s(T)$ to the mode coupling form
$\tau_s(T)\sim (T-T_{C})^{-\gamma}$ to find $T_C$. For the
self part of the intermediate scattering function, the actual
value of $\tau_{s}$ increases as the magnitude of the wave vector
decreases \cite{Kob95b}. However, the value of $T_C$ is independent
of $k$. $\tau_s(T)$ versus temperature and the mode coupling fit are 
shown in Figure \ref{fig:mct_tau}. We find the best fit with the mode 
coupling temperature $T_C = 0.303$ which corresponds to $\Gamma=1.46$.
Note that $T_C$ is determined from measurements made at temperatures where
the system is equilibrated.
Also shown in Figure \ref{fig:mct_tau} is the fit to the Vogel--Fulcher
form $\tau_s(T)=A\exp[B/(T-T_{VF})]$ with $T_{VF}=0.21$ which corresponds
to $\Gamma=1.60$. In doing
the Vogel--Fulcher fit, we were able to use a much broader range
of temperatures since $T_{VF}$ is much lower than the mode coupling
$T_{C}$.
\begin{Fig}{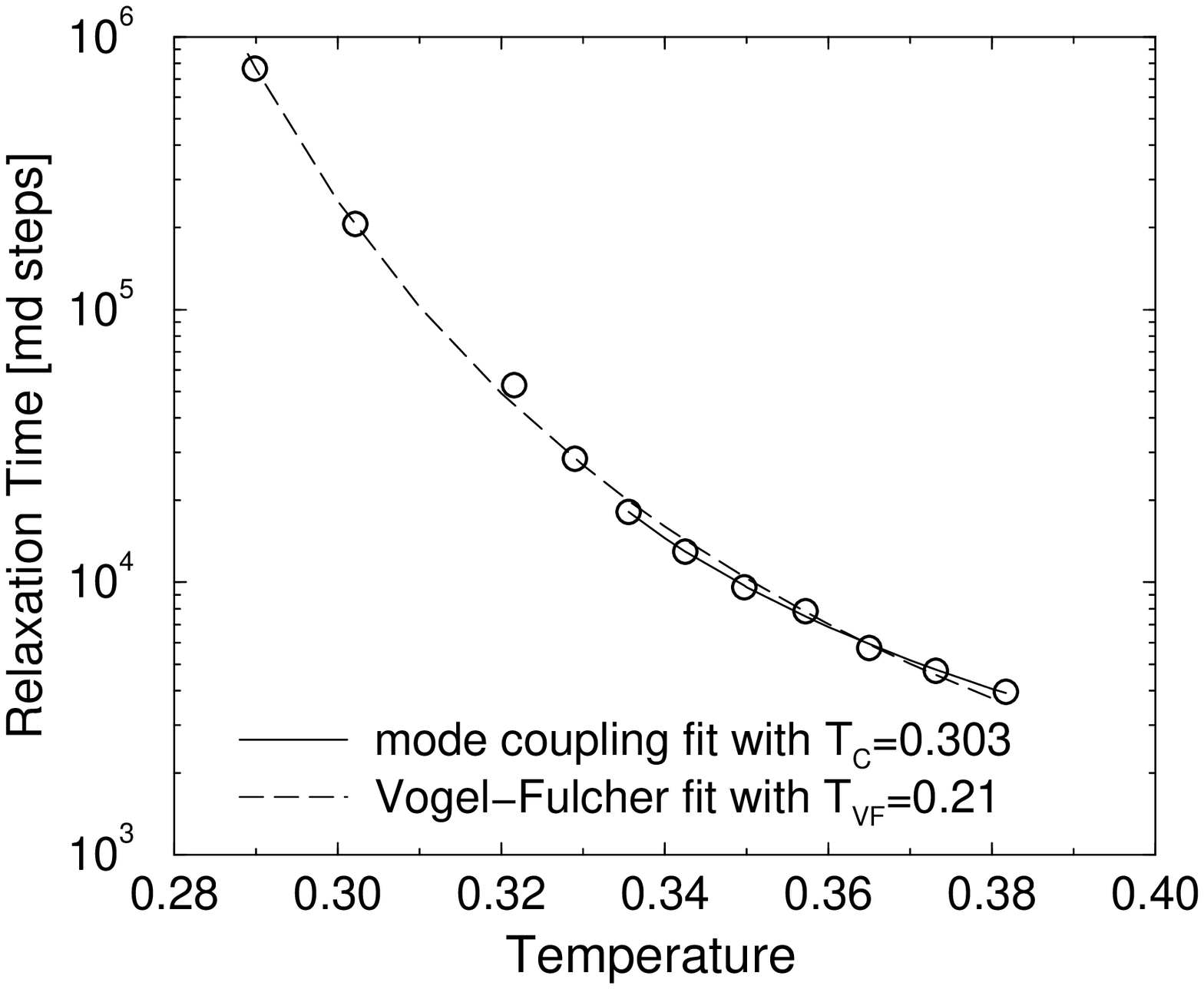}
\caption{Relaxation times $\tau_s$ versus temperature. The solid line
is the mode coupling fit to the form $\tau_s=A(T-T_{C})^{-\gamma}$
with $T_C = 0.303$, $\gamma=1.735$ and $A=47.6$. The dashed line is the fit to
the Vogel--Fulcher form $\tau_s(T)=A\exp[B/(T-T_{VF})]$ with $T_{VF}=0.21$,
$A=33.3$, and $B=0.803$.}
\label{fig:mct_tau}
\end{Fig}

The full intermediate scattering function $F(\vec{k},t)$ is given
by \cite{Hansen}
\begin{equation}
F_{\alpha}(\vec{k},t)=\frac{1}{N_{\alpha}}\left\langle
\rho_{\vec{k},\alpha}(t)\rho_{-\vec{k},\alpha}(0)\right\rangle
\label{eq:fullFkt}
\end{equation}
where the Fourier transform of the density
$\rho_{\vec{k}}(t)=\sum_{i=1}^{N}e^{-i\vec{k}\cdot\vec{r}_{i}(t)}$.
The subscript $\alpha$ refers to the particle type, A or B.
The longest $\alpha$ relaxation time can be determined from
the full intermediate scattering function evaluated at $k=k_{max}$
\cite{Rinaldi01}. We set $k_{max}=2\pi \cdot 8.3666/L$ ($L=8$) which is 
the location of 
the first peak in the structure factor for type B particles.
We define the $\alpha$ relaxation time $\tau$ as the time where
$F_{\alpha}(k_{max},t)$ decays to $1/e$. At a temperature 
$T=0.2895858$ which is just below the mode coupling $T_C$,
we find that $F_B(k,t)$ has fallen to $1/e$ at 
$\tau=(1.0\pm 0.1)\times 10^{6}$ md time steps for a system with
512 particles of which half are type B. This gives us a time
scale by which to compare other times such as our run times. 
This value of $\tau$ shows no signs of aging \cite{Kob97b,Kob00}
and stays about the same even after 10$^8$ time steps.
At higher temperatures this relaxation time is much shorter.

The runs used to determine the intermediate scattering function
were done in a slightly different way from the other measurements.
These runs were performed at a given temperature and density for 
$N_{\tau}$ md time steps with no change in temperature or density.
The runs were started from a configuration that had been equilibrated 
at that temperature and density using parallel tempering. The parallel
tempering technique is described in the appendix.  

\section{Specific Heat}
The specific heat is a thermodynamic quantity which undergoes a change
signaling the glass transition. 
In experimental systems under constant pressure the specific heat
exhibits a smooth step down as the temperature is lowered through the
glass transition. In our simulations which are done at constant volume,
the specific heat has a peak at the glass transition. It is a useful 
check of our calculation to see if the peak occurs at the same temperature
(or density) as the drop in the linear generalized compressibility. 
There are two ways
to compute the specific heat $C_V$ per particle at constant volume $V$. 
The first is by taking a derivative
of the average energy $\langle E\rangle$ per particle with respect to 
temperature: $C_V=d\langle E\rangle/dT$. Since
we study the system at discrete temperatures, we approximate the
derivative by a finite difference:
\begin{equation}
C_V(T_n)=\frac{\langle E(T_n)\rangle -\langle E(T_{n-1})\rangle}{T_n-T_{n-1}}
\label{eq:sphtDeriv}
\end{equation}
where $T_n>T_{n-1}$ for all integers $n$. The second way to calculate
the specific heat is from the fluctuations:
\begin{equation}
C_V=N k_B\beta^2\left(\langle E_P^2\rangle - \langle E_P\rangle^2\right)
\end{equation}
where $k_B$ is Boltzmann's constant, $\beta$ is the inverse temperature,
and $E_P$ is the potential energy per particle.
In our three dimensional simulations the kinetic energy per
particle is given by 
$3k_BT/2$, so it is the fluctuations in the potential energy $E_P$ 
per particle which
determine the temperature dependence of the specific heat. Thus
\begin{equation}
C_V=\frac{3}{2}k_B+
Nk_B\beta^2\left(\langle E_P^2\rangle - \langle E_P\rangle^2\right)
\label{eq:sphtFluct}
\end{equation}
In equilibrium these two ways of calculating the specific heat should
agree. So we compare the results of calculating $C_V$ both ways
as a check on our calculation and to make sure the system has equilibrated
in all the basins that were visited in the energy landscape. 

\subsection{Specific Heat Versus Temperature}
The specific heat at constant volume exhibits a peak at the glass
transition as shown in Figure \ref{fig:sphtFluctdEdT}. The data in this
figure is for 512 particles and was averaged over 6 runs
with a measurement time of $3\times 10^6$ md steps. Notice that there
is good agreement between calculating the specific heat by taking a derivative
of the energy with respect to temperature (see eq. (\ref{eq:sphtDeriv}))
and by using fluctuations (see eq. (\ref{eq:sphtFluct})). This implies that
the system has equilibrated within the basins that it visits
in the energy landscape. We find similar agreement for other run
times. At low temperatures the
specific heat goes to 3$k_B$ as expected for classical oscillators while
at high temperatures $C_V$ approaches $3k_B/2$ which corresponds to an
ideal gas. 
\begin{Fig}{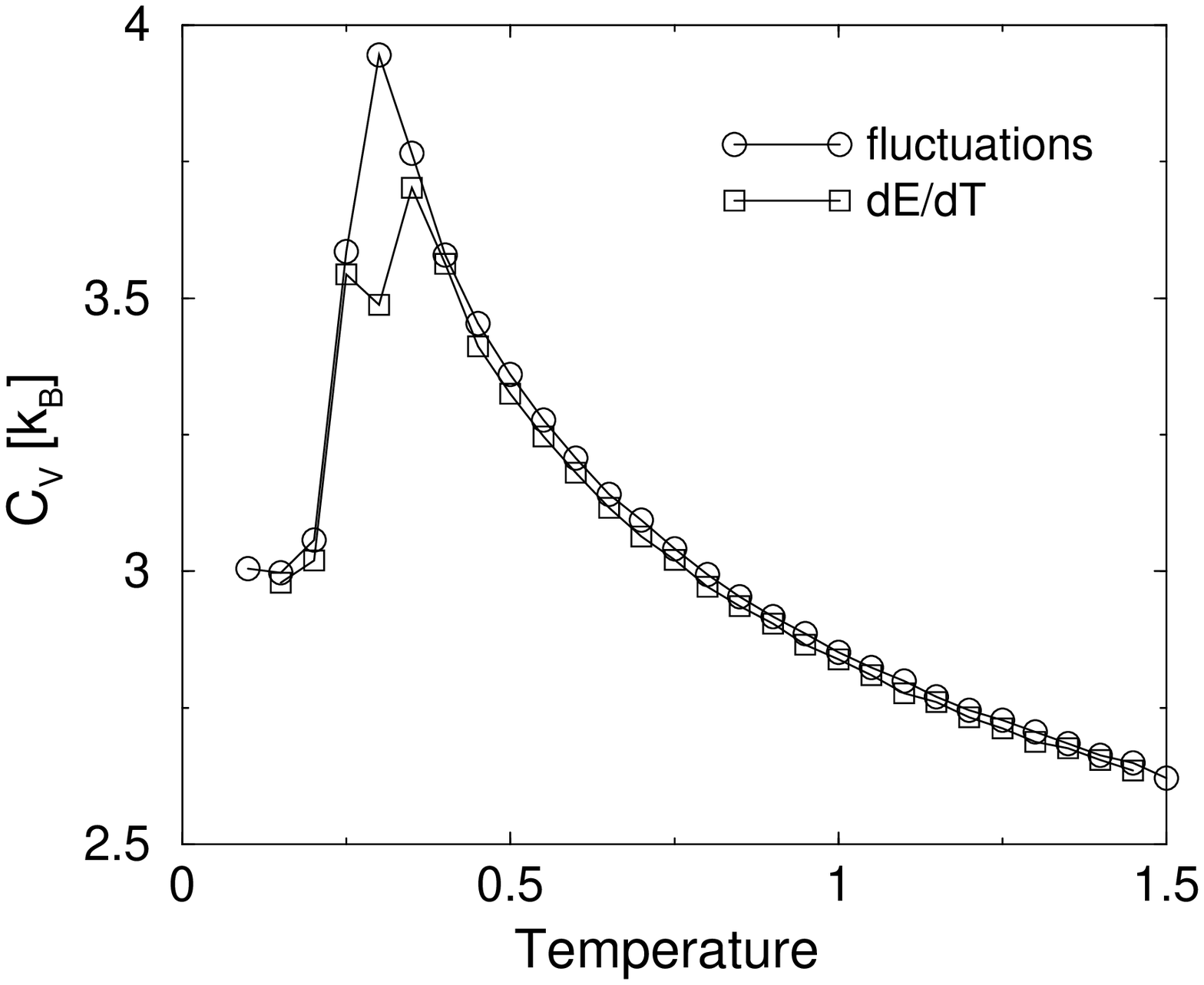}
\caption{Specific heat at constant volume as a function of
temperature for binary mixture of 512 particles with a measuring time of 
$3\times 10^6$ md steps averaged over 6 runs.
$\rho_o = 0.6$ and $\sigma_B/\sigma_A=1.4$. The specific heat is calculated
from energy fluctuations and by taking the derivative of the energy
with respect to temperature.}
\label{fig:sphtFluctdEdT}
\end{Fig}
The peak in the specific heat occurs at $T\approx 0.3$ which corresponds
to $\Gamma\approx 1.46$. The temperature of the peak coincides with
the mode coupling $T_C=0.303$ that we deduced from the intermediate
scattering function data.
Longer run times lead to a sharper peak in the specific heat as can
be seen in Figure \ref{fig:spht_8_time} which shows the specific heat
for 512 particles
for several different measuring times. The peaks would presumably be
sharper if we had used a finer temperature scale. At high temperatures
the agreement between the different times is very good. Perera
and Harrowell \cite{Perera99} have found a specific heat peak
in a two dimensional binary mixture of soft spheres. They argue
that their peak is an equilibrium feature. However, in our
case, at temperatures below the peak, the system
has fallen out of equilibrium and has become trapped in a basin
in the energy landscape. We shall see this later by examining the 
energy of the inherent structures
(potential energy minima) as a function of temperature.
Thus the fact that the peak in the specific heat occurs at or very close
to the mode coupling $T_C$ is a result of the 
relaxation times (see Fig. \ref{fig:mct_tau}) becoming comparable to 
and exceeding the simulation run times as the temperature drops below $T_C$.
When this happens, the system falls out of equilibrium and undergoes
a kinetic glass transition.
\begin{Fig}{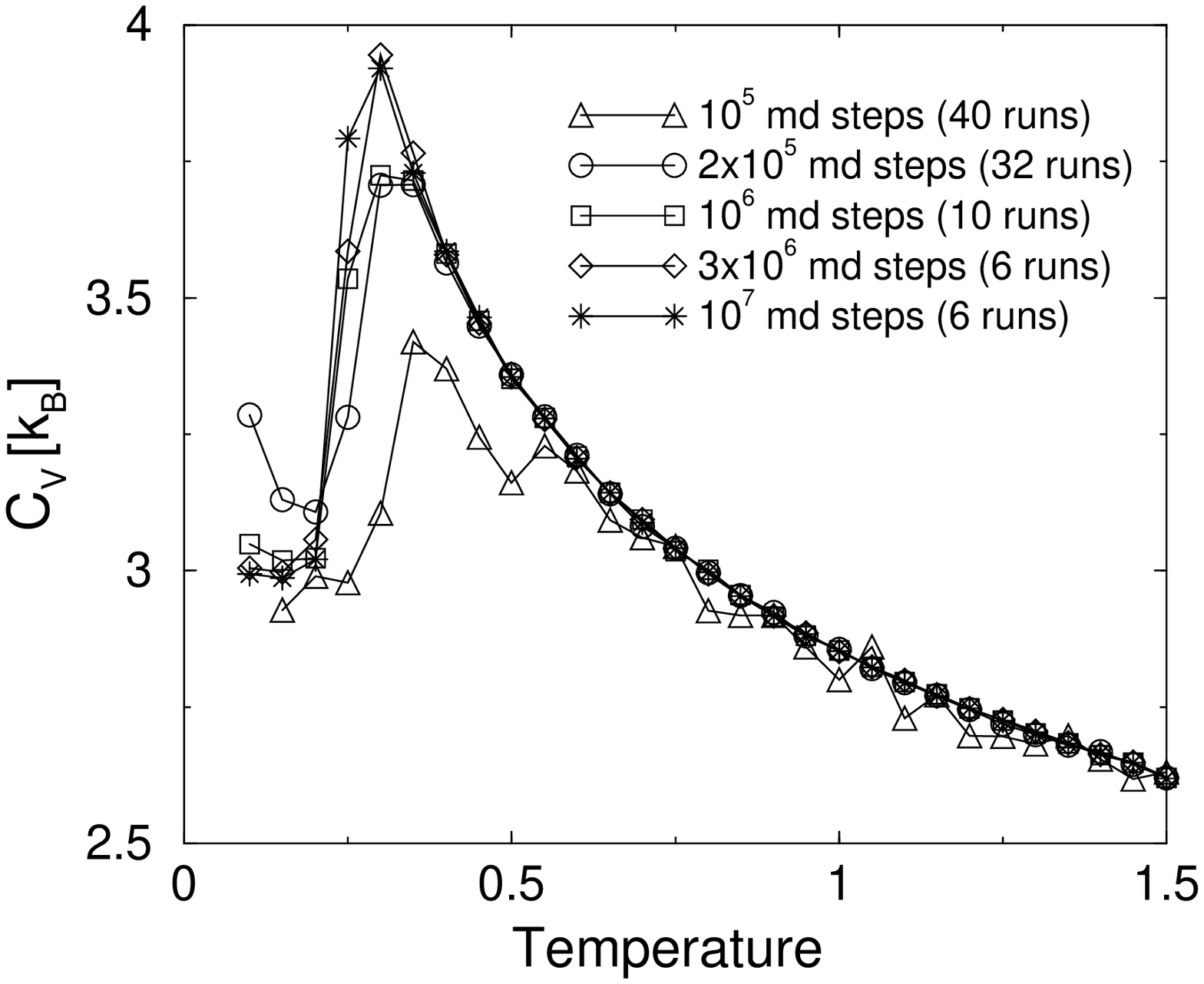}
\caption{Specific heat at constant volume as a function of
temperature for binary mixture of 512 particles with measuring times of $10^5$,
$2\times 10^5$, $10^6$, $3\times 10^6$, and $10^7$ md time steps.
The number of runs averaged over is indicated in the legend. The specific
heat is calculated from energy fluctuations. $\rho_o = 0.6$ and 
$\sigma_B/\sigma_A=1.4$.}
\label{fig:spht_8_time}
\end{Fig}

The specific heat $C_P$ of experimental systems at constant pressure
exhibits a downward step at the glass transition during cooling and a 
peak at slightly higher temperatures upon heating \cite{DeBolt76}. 
As can be seen in Figure
\ref{fig:sphtCoolHeat}, in our warming up simulations,
which are done at constant volume, the specific heat peak sharpens and
moves toward higher temperatures compared to the cooling runs.
This is consistent with what is seen in experiments. 
The hysteresis is consistent with the system falling out of
equilibrium and getting stuck in a basin of the energy landscape.
\begin{Fig}{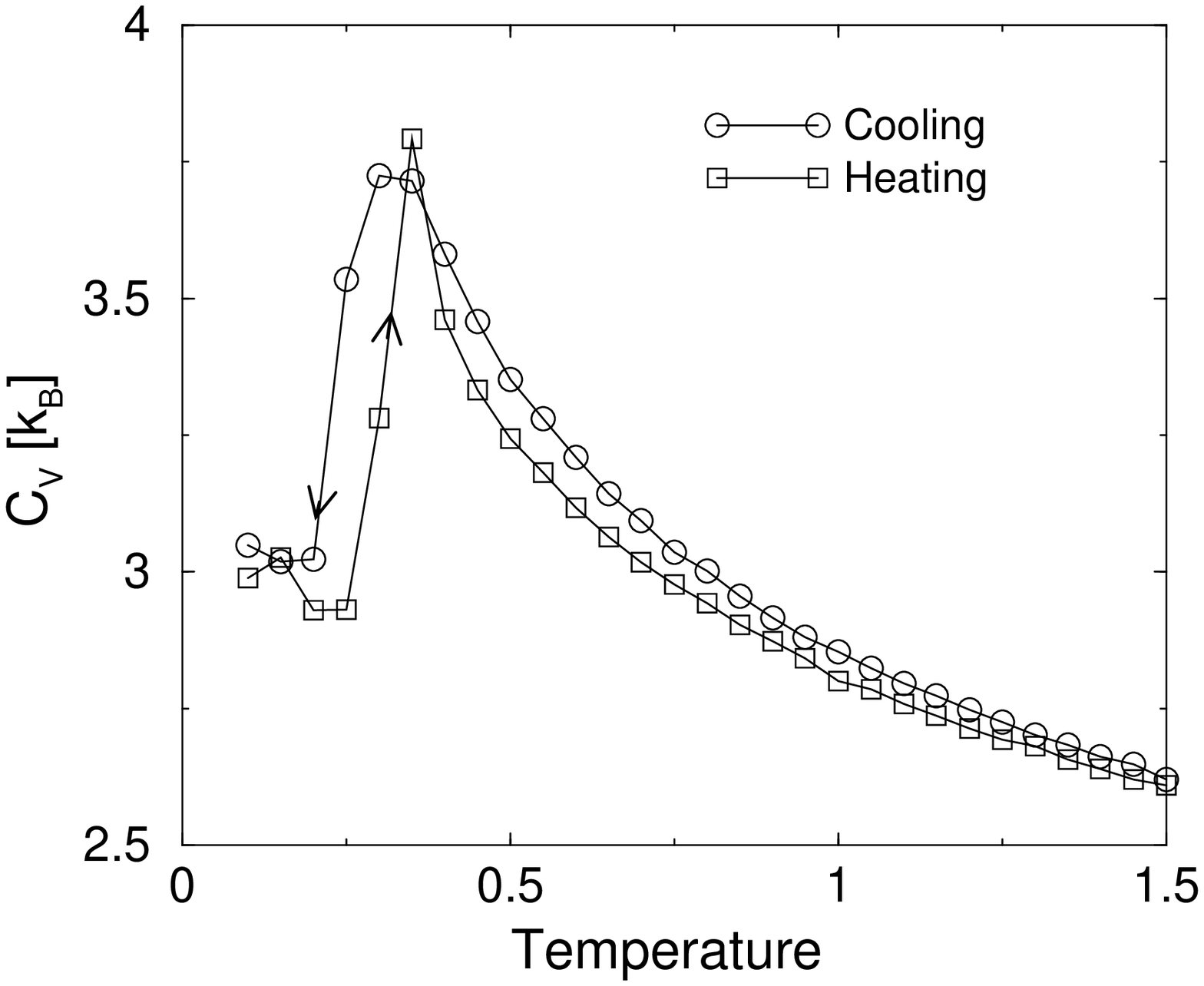}
\caption{Specific heat at constant volume during heating and cooling a
binary mixture of 512 particles with a measuring time of $10^6$ md time steps
averaged over 10 runs. $\rho_o = 0.6$ and $\sigma_B/\sigma_A=1.4$.}
\label{fig:sphtCoolHeat}
\end{Fig}

As we mentioned in the introduction, some have suggested that the
glass transition has an underlying second order phase transition
\cite{Kirkpatrick89,Sethna91,Kivelson95,Ernst91,Dasgupta91,Menon95}.
Unlike typical second order phase transitions, there is no experimental 
evidence that the specific heat diverges at the glass transition. This is
consistent with our simulations. In simulations one looks for a divergence
by examining whether the quantity increases systematically with system
size. In Figure \ref{fig:sphtSize} we plot $C_V$ for systems with 64, 216, 
512 and 1000 particles. As one can see, the specific heat does not exhibit any
size dependence. However, we cannot rule out the possibility that a
thermodynamic phase transition occurs at temperatures below where 
we fall out of
equilibrium. Indeed theories which postulate a thermodynamic transition
put the transition temperature well below the mode coupling $T_C$.
\begin{Fig}{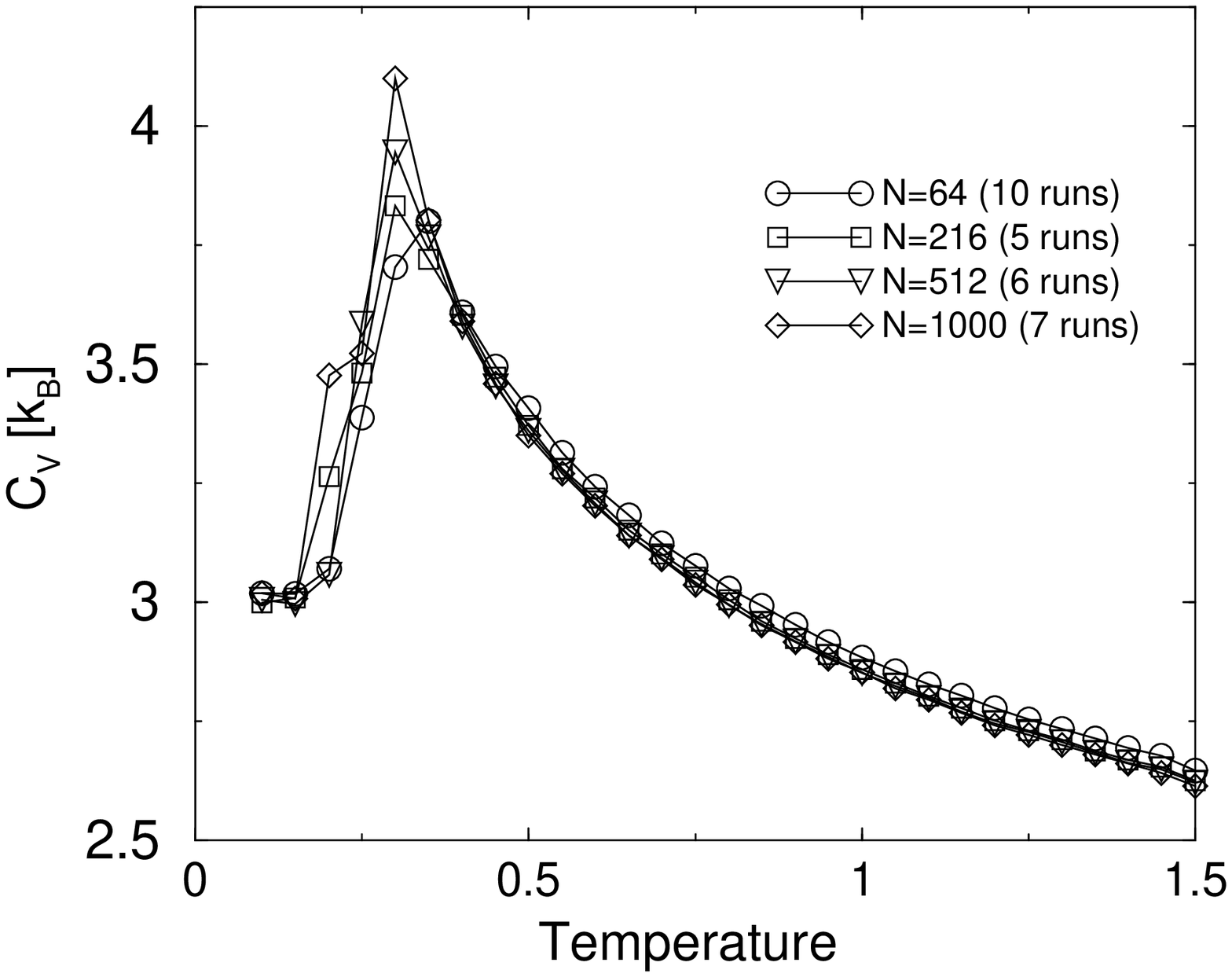}
\caption{Specific heat during cooling for binary mixtures of 64, 216, 
512, and 1000 particles. The measuring time was $3\times 10^6$ md time 
steps. The specific heat was calculated from fluctuations and averaged 
over the number
of runs indicated in the legend. Note the lack of size dependence.
$\rho_o = 0.6$ and $\sigma_B/\sigma_A=1.4$.} 
\label{fig:sphtSize}
\end{Fig}

\subsection{Specific Heat Versus Density}
In Figure \ref{fig:sphtdens} we show the specific heat as a function
of density. As the density increases, the specific heat rises to
a peak at $\rho_{o}^{\rm peak}=0.8$. This corresponds to
$\Gamma=1.44$ which is in good agreement with the $\Gamma$
value of 1.46 that we found for the specific heat peak when we varied
the temperature. Going to higher densities corresponds to going to
lower temperatures.
At densities higher than 0.8, the system falls out of equilibrium.
\begin{Fig}{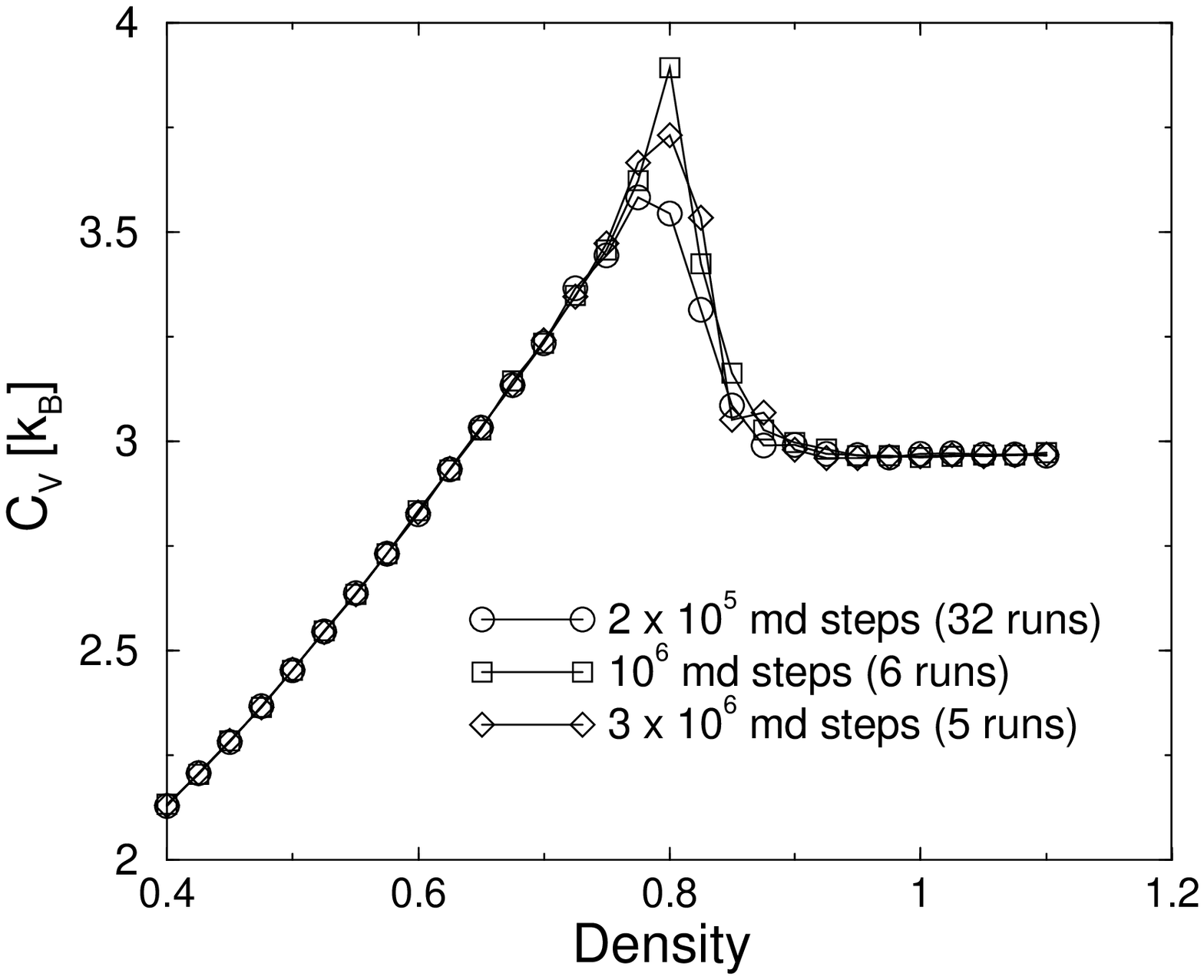}
\caption{Specific heat versus density for a binary mixture of 512
particles with $T=1$. The measuring times were $2\times 10^5$,
$10^6$ and $3\times 10^6$ md time steps. The specific
heat was calculated from fluctuations and averaged over the number
of runs indicated in the legend. 
$\rho_o = 0.6$ and $\sigma_B/\sigma_A=1.4$.} 
\label{fig:sphtdens}
\end{Fig}

\section{Generalized Compressibilities}
As we mentioned in the introduction, the generalized compressibilities
are thermodynamic probes that are a function of the microscopic structure
of the system. They are solely a function of the positions of the particles
and do not depend on their histories. So one could take snapshots of the
configurations of the particles at different instances, scramble the order
of the snapshots, and still be able to calculate the generalized
compressibilities. Averaging over these snapshots corresponds to ensemble
averaging. In this sense the generalized compressibilities are
thermodynamic quantities which can be calculated solely from the microstates
of the system and do not depend on the system's dynamics or kinetics.

We now derive expressions for
the linear and nonlinear generalized compressibility. 
To probe the density fluctuations, we follow the approach
of linear response theory and consider applying an external 
potential ${\Delta P\over\rho_o}\phi(\vec r)$ which couples to the
local density $\rho (\vec r)=\sum_{i=1}^{N}\delta (\vec r - \vec r_i)$
where $\vec r_i$ denotes the position of the $i^{th}$ particle.
$\rho_o$ is the average density.
$\Delta P$ has units of pressure and sets the magnitude of the perturbation.
$\phi (\vec r)$ is a dimensionless function of
position that must be compatible with the periodic boundary conditions
imposed on the system, i.e., it must be continuous across the
boundaries, but is otherwise arbitrary.
This adds to the Hamiltonian H of the system a term
\begin{equation}
U={\Delta P\over\rho_o}\int_Vd^3r\phi (\vec r)\rho (\vec r)
={\Delta P\over\rho_o}\sum_i\phi (\vec r_i)
\equiv {\Delta P\over\rho_o}\rho_{\phi}
\label{eq:U}
\end{equation}
where we have defined 
$\rho_{\phi}=\int_Vd^3r\phi (\vec r)\rho (\vec r)=\sum_i\phi (\vec r_i)$.
$\rho_{\phi}$
is the inner product of $\phi$ and $\rho(\vec r)$, and we can regard
it as a projection of the density onto a basis function $\phi(r)$,
i.e., $\rho_{\phi}=<\rho|\phi>$. 
It weights the density fluctuations according to their spatial position. 
The application of the external potential will induce an average change 
$\delta \rho_{\phi}$ in $\rho_{\phi}$: 
\begin{equation}
\delta \rho_{\phi}=\langle\rho_{\phi}\rangle_{U}-
\langle\rho_{\phi}\rangle_{U=0}
\label{eq:deltarho}
\end{equation}
where the thermal average $\langle\rho_{\phi}\rangle_{U}$ is given by
\begin{equation}
\langle \rho_{\phi}\rangle_{U}={1\over Z}\hbox{\rm Tr}
\left[ e^{-\beta (H+U)}\rho_{\phi}\right]\quad 
\label{eq:AV}
\end{equation} 
The partition function $Z=\hbox{\rm Tr}e^{-\beta (H+U)}$ and
$\beta$ is the inverse temperature.
For small values of $\Delta P$, this
change can be calculated using perturbation theory \cite{Toda91}.
Up to third order in $\Delta P$, we find
\begin{eqnarray}
\label{eq:expansion}
\nonumber
\delta \rho_{\phi}&=& 
        -\frac{\beta\Delta P}{\rho_o}\langle \rho_{\phi}^2\rangle_c +
        {\beta^2\Delta P^2\over 2\rho_o^2}\langle 
        \rho_{\phi}^3\rangle_c\\ 
        && -{\beta^3\Delta P^3\over 6\rho_o^3}
        \langle \rho_{\phi}^4\rangle_c,
\end{eqnarray}
where the cumulant averages are 
\begin{eqnarray}
\langle \rho_{\phi}^2\rangle_c&=&\langle \rho_{\phi}^2\rangle -
        \langle \rho_{\phi}\rangle^2 
\label{eq:CUMU2}\\
\langle \rho_{\phi}^3\rangle_c&=&\langle \rho_{\phi}^3\rangle -
        3\langle \rho_{\phi}\rangle\langle \rho_{\phi}^2\rangle +
        2\langle \rho_{\phi}\rangle^3 \label{eq:CUMU3}\\ \nonumber
\langle \rho_{\phi}^4\rangle_c&=&\langle \rho_{\phi}^4\rangle -
        4\langle \rho_{\phi}\rangle\langle \rho_{\phi}^3\rangle -
        3\langle \rho_{\phi}^2\rangle^2 + \\
        && 12\langle \rho_{\phi}\rangle^2\langle \rho_{\phi}^2\rangle -
        6\langle \rho_{\phi}\rangle^4
\label{eq:CUMU4}
\end{eqnarray}
with the thermal average
$\langle \rho_{\phi}^n\rangle=\langle \rho_{\phi}^n\rangle_{U=0}$.
The third order cumulant, eq.(\ref{eq:CUMU3}), is zero in the liquid phase
because for every configuration there exists
an equivalent configuration with the opposite sign 
of $\left(\rho_{\phi}-\langle \rho_{\phi}\rangle\right)$ and so we will not 
consider this term any further. 
We can recast eq. (\ref{eq:expansion}) as a power
series in the perturbation $\Delta P$:
\begin{equation}
\label{eq:DV2}
{\delta\rho_{\phi} \over {N}}=
-{1\over 6\rho_o k_B T}\chi_{l}\Delta P+{1\over 6(\rho_o k_B T)^3}\chi_{nl}
(\Delta P)^3
\end{equation}
where 
\begin{equation}
\chi_{l}={6 \over N}\langle (\rho_{\phi})^2\rangle_c \quad\;\;\;\;
\chi_{nl}=-{ 1 \over N}\langle (\rho_{\phi})^4\rangle_c.
\label{eq:susceptibilities}
\end{equation}
In this paper we
will focus our attention on the linear ($\chi_{l}$) and nonlinear
($\chi_{nl}$)  dimensionless generalized compressibilities defined by the
above expressions. 

We now discuss the choice of the function $\phi$.
We consider applying the potential along the
direction $\mu$ of one of the coordinate axes so that $\phi(\vec
r)=\phi(r^{\mu})$. A
natural candidate for $\phi(r^{\mu})$ is $\cos(k_{\mu}r^{\mu})$ (no implied
sum over repeated indices) with $k_{\mu}=2\pi n/L$, where $n=1,2,...$ 
In this case, $\rho_{\phi}$ is the $k^{th}$ mode of
the cosine transform of the density. We will also consider the
simpler function $\phi(r^{\mu})=|r^{\mu}|/L$. The absolute value
corresponds to the case where all the particles feel a force along the
$\mu$th direction pointing towards the origin. It
gives results very similar to $\phi(r^{\mu})=\cos(k_{\mu}r^{\mu})$ for
small $k$ at a fraction of the computational cost. (No sum over
repeated indices.) So our results in
this paper correspond to two cases:
\begin{equation}
\rho_{\phi}=\sum_{i}|r^{\mu}_{i}|/L
\label{eq:abs}
\end{equation}
which is rather like a center of mass, 
and 
\begin{equation}\rho_{\phi}=\sum_{i}\cos(k_{\mu}r_{i}^{\mu})
\label{eq:cosine}
\end{equation}  
Since the system is isotropic, we compute the compressibilities
for each direction and then average over the direction $\mu$.

In most of our calculations we work in the canonical ensemble
where we fix the volume $V$, the number $N$
of particles and the density $\rho_o$. However, it is straightforward
to generalize our results to the grand canonical ensemble
where the number of particles is not fixed. We simply replace the
thermal average defined in equation (\ref{eq:AV}) by
\begin{equation}
\langle \rho_{\phi}\rangle_{U}={1\over {\cal Z}}\sum_{N}
e^{\mu N}\hbox{\rm Tr}
\left[ e^{-\beta (H_{N}+U_{N})}\rho_{\phi}\right]\quad
\label{eq:grandcanonicalAV}
\end{equation}
where $\mu$ is the chemical potential, $H_N$ is the Hamiltonian with
$N$ particles, $U_N$ is given by eq. (\ref{eq:U}) for a system with $N$
particles, and ${\cal Z}$ is the grand canonical partition function
given by
\begin{equation}
{\cal Z}=\sum_{N}e^{\mu N}\hbox{\rm Tr}
\left[e^{-\beta (H_{N}+U_{N})}\right]
\end{equation}
The generalized compressibilities can be defined
using equations (\ref{eq:CUMU2}) through (\ref{eq:susceptibilities})
with the thermal averages
$\langle \rho_{\phi}^n\rangle=\langle \rho_{\phi}^n\rangle_{U=0}$
defined in the grand canonical ensemble.

\subsection{Results for Linear Generalized Compressibility}
We now turn to our results for the binary glass forming liquid. 
\subsubsection{$\chi_{l}$ from Absolute Value of Positions versus Temperature}
We will first discuss the linear generalized compressibility calculated
from the absolute values of the particle positions using 
eqs. (\ref{eq:susceptibilities}) and (\ref{eq:abs}).  
\begin{Fig}{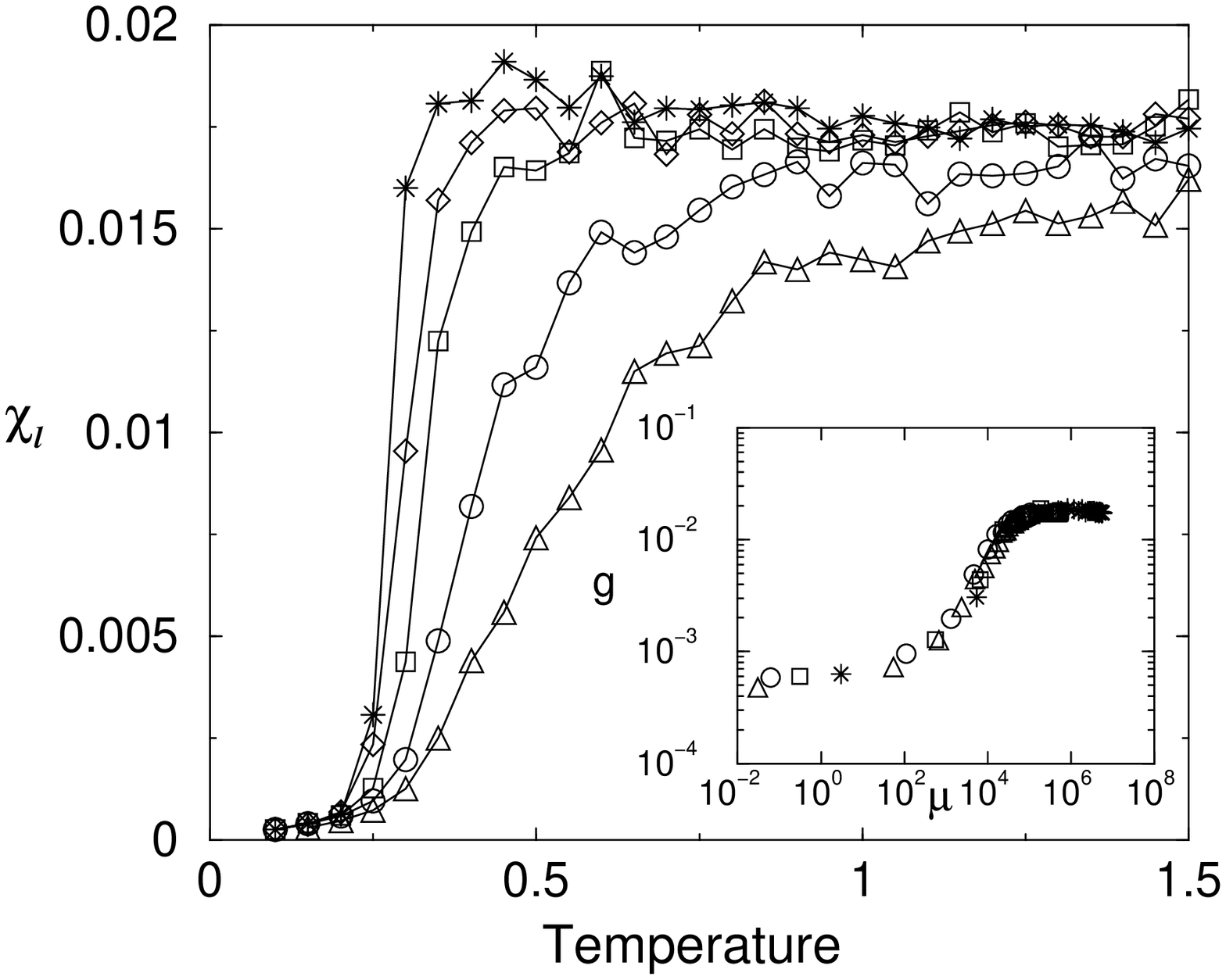}
\caption{Linear generalized compressibility as a function of
temperature for different measuring times $t_M$: $10^5$ ($\triangle$, 40
runs), $2\times 10^5$ ($\circ$, 32 runs), $10^6$ ($\Box$, 10 runs), 
$3\times 10^6$ ($\Diamond$, 6 runs) and $10^7$ ($\ast$, 6 runs) 
md steps. System size is 512
particles. $\rho_o = 0.6$ and $\sigma_B/\sigma_A=1.4$. $\chi_{l}$ is
calculated using the absolute value of the particles' positions. Inset: 
$T>T_o$ subset of the same data scaled as described in text.}
\label{fig:glass_lin_susc}
\end{Fig}
Figure \ref{fig:glass_lin_susc} shows the linear generalized
compressibility as a function of temperature for different run times.
The compressibility at high temperatures is independent of $T$. 
In the vicinity of the glass transition $\chi_{l}$
drops. Notice that as the measuring time $t_M$ increases (and
hence as the cooling rate decreases), the temperature
of the drop decreases and becomes more abrupt. The measuring time
can be thought of as the number of snapshots at a single temperature that we 
use to calculate the compressibility.
The linear compressibility is proportional to the width of the
distribution of $\rho_{\phi}$, so the drop in $\chi_{l}$ corresponds to
the sudden narrowing of the distribution $P(\rho_{\phi})$. 
If we regard $\rho_{\phi}$ as a generalized center of mass, then 
the drop in $\chi_{l}$ signals the sudden arrest in the fluctuations 
of the generalized center of mass. In other words,
at the glass transition the motion of the particles is largely frozen and
hence, the generalized center of mass does not move around much. This is
consistent with recent observations of the colloidal glass transition
in which the size of the clusters of ``fast'' particles drops dramatically
at the glass transition \cite{Weeks00}.

Notice that at longer measuring times, the temperature $T_{\rm drop}$
at which the generalized linear compressibility drops is roughly at
the mode coupling temperature $T_C=0.303$. Let us define 
$T_{\rm drop}$ as the temperature at which $\chi_{l}$ has dropped
halfway down. For $10^6$ md steps, 
$T_{\rm drop}\approx 0.33$; for $3\times 10^6$ md steps,
$T_{\rm drop}\approx 0.30$; and for $10^7$ md steps,
$T_{\rm drop}\approx 0.27$. Thus we are able to stay in equilibrium
down to the mode coupling temperature for our longer runs. This is
what we would expect when we
compare these run times, which are longer than 1 million time steps,
to the $\alpha$ relaxation time $\tau$ which is about
1 million time steps at $T=0.29$ which is just below $T_C$. 
Thus the fact that the drop in the linear generalized compressibility
occurs at or very close to the mode coupling $T_C$ is a result of the
relaxation times (see Fig. \ref{fig:mct_tau}) becoming comparable to 
and exceeding the simulation run times as the temperature drops below $T_C$.
When this happens, the system falls out of equilibrium and undergoes
a kinetic glass transition.

The behavior exhibited by $\chi_l$ can be quantified using a scaling ansatz:
$\chi_{l}(t_M,T)=g(\mu=t_M/\tau (T))$, where the characteristic time
has the Volgel--Fulcher form $\tau(T)=\exp (A/(T-T_o))$. The inset
of Figure \ref{fig:glass_lin_susc} shows that the data collapse onto
a single curve with $A=0.75$, $T_o=0.15$. (The data could not be
fitted using $\tau (T)=A(T-T_o)^{\gamma}$ as suggested by simple mode
coupling theories \cite{Gotze92}.) Notice that $T_o$ lies below
the Vogel--Fulcher temperature $T_{VF}=0.21$ and the MCT critical temperature
$T_{C}=0.303$ deduced by fitting the temperature dependence of the 
relaxation times. This scaling suggests that
$\chi_{l}$ becomes a step function for infinite $t_M$ and that
the drop in the
compressibility would become a discontinuity at infinitely long times.
This abrupt drop is consistent with a sudden arrest of the motion of the
particles in the liquid which is the kinetic view of the glass
transition. The abrupt drop also appears to be
in agreement with Mezard and Parisi's proposal that the glass
transition is a first order phase transition 
with a jump in the specific heat \cite{Mezard99,Parisi97b}. 
Indeed the temperature at which $\chi_{l}$ drops agrees with the
temperature of the peak in the specific heat shown in 
Figure \ref{fig:spht_8_time}. 
The specific heat provides an independent check of the 
glass transition temperature.
However, the drop in $\chi_l$ and the specific heat peak
are due to the system falling out
of equilibrium, and therefore we cannot really tell if there is an 
underlying true thermodynamic transition. 

One way to see that the system is falling
out of equilibrium is to plot the inherent structure energy per particle
\cite{Stillinger82,Stillinger84,Sastry98}.
An inherent structure is a particular system configuration
whose energy corresponds to the minimum of a basin in the energy
landscape. The energy landscape is a 3N dimensional surface defined
by the potential energy of the system which is a function of the
particles' coordinates. During each run we sampled the 
configurations found at each temperature. Each configuration lies
somewhere in a basin and we used the method of conjugate gradients
\cite{Press92} to find the inherent structure energy of that basin.
The result is shown in Figure \ref{fig:inherent_structure} where we
plot the average inherent structure energies versus the temperature
of the configuration that was originally saved. At
high temperatures the inherent structure energy $e_{IS}$ per
particle is flat as a function of temperature. As the system
is cooled, $e_{IS}$ decreases rather steeply \cite{Sastry98}.
The inherent
structure energy flattens off at low temperatures where the system has
fallen out of equilibrium and has become stuck in one basin. 
For each measuring time the 
temperature below which the generalized linear compressibility
drops corresponds to the temperature below which the inherent
structure energy flattens off at low temperatures. Thus the temperature of
the drop in $\chi_l$ and the peak in the specific heat corresponds 
to the temperature below which the system falls out of equilibrium
and ceases to explore deeper basins of the energy landscape.
\begin{Fig}{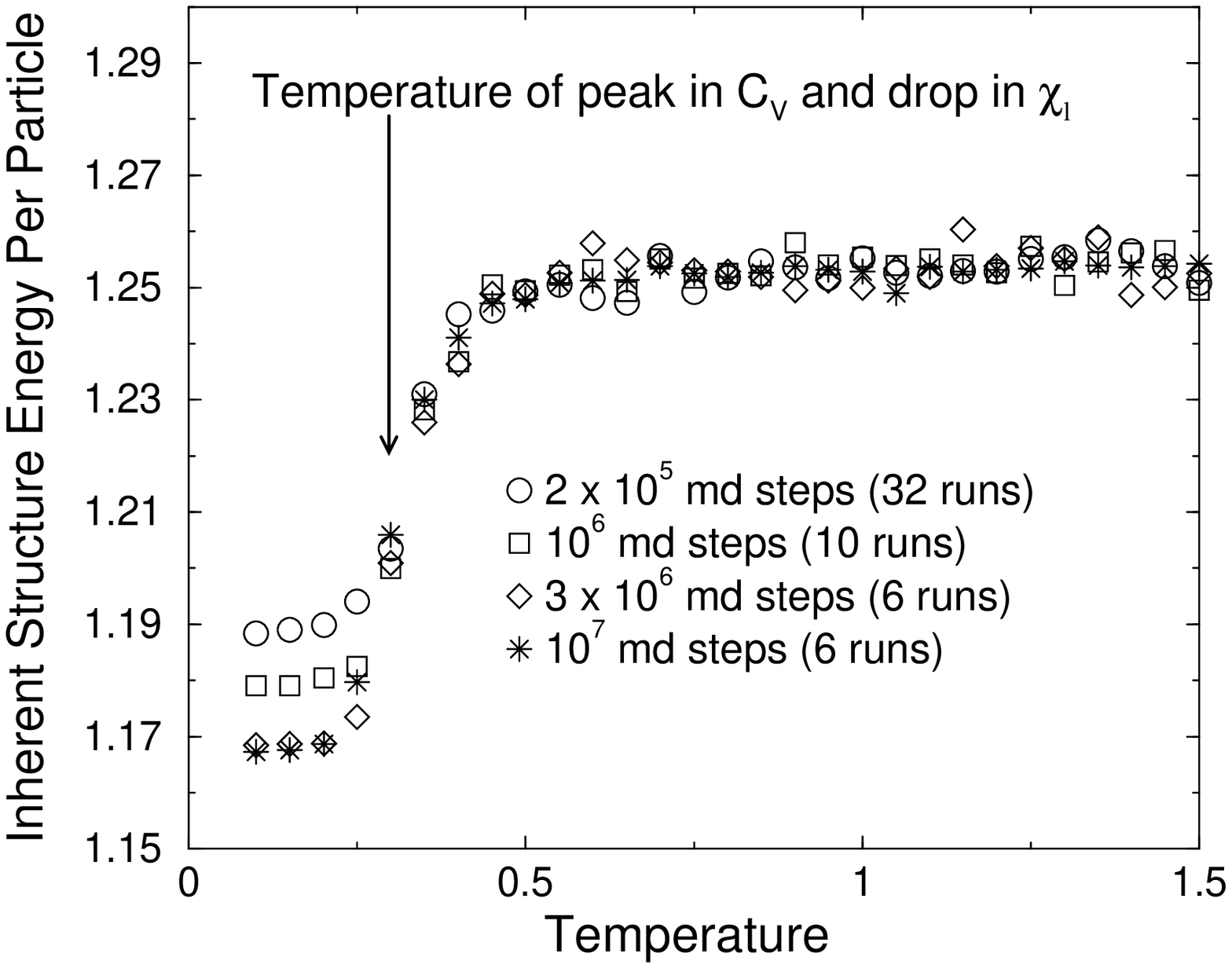}
\caption{Inherent structure energy per particle
as a function of temperature for a system of 512 particles 
at different measuring times. Other parameters are the same as
in Figure \protect\ref{fig:glass_lin_susc}.}
\label{fig:inherent_structure}
\end{Fig}

\begin{Fig}{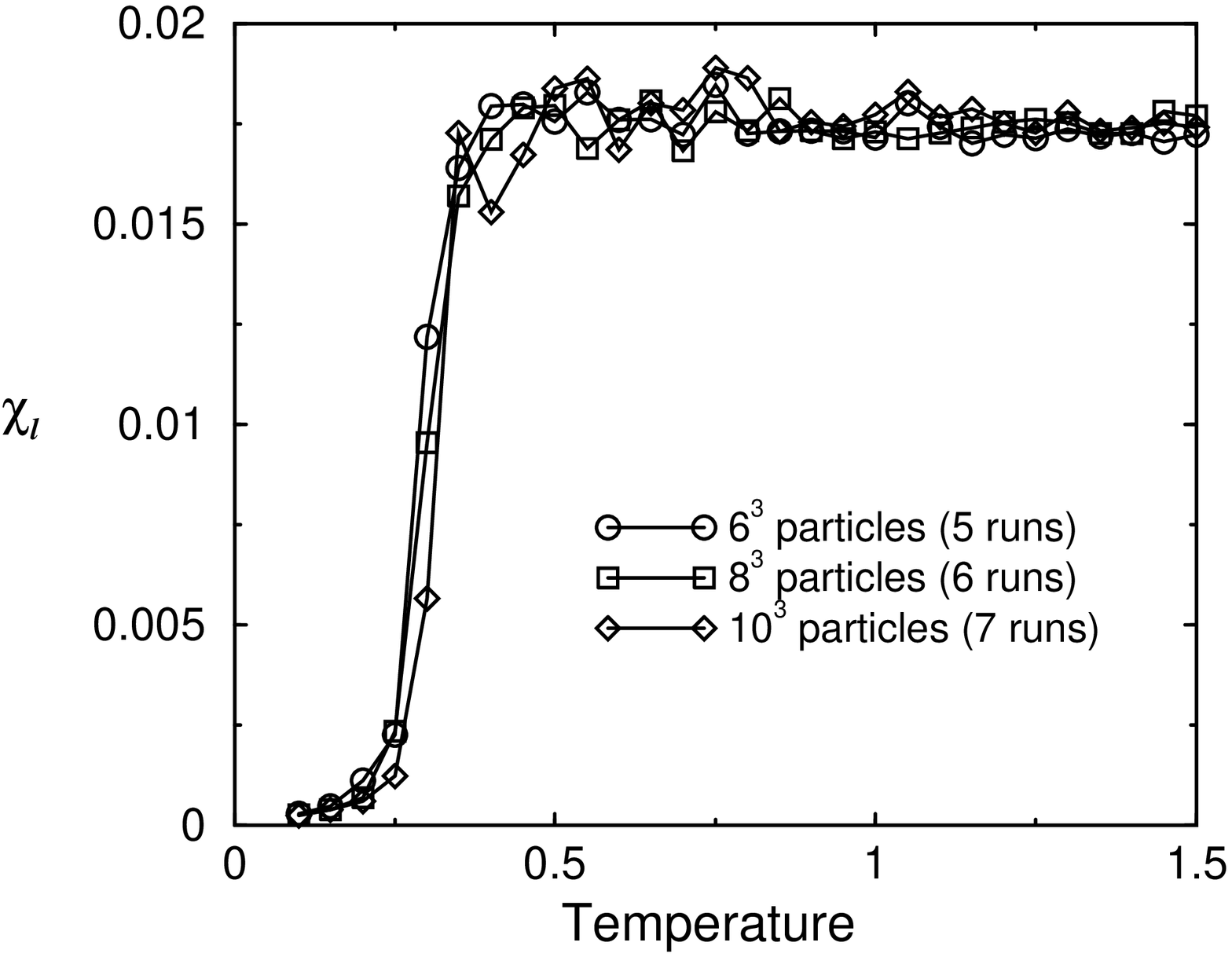}
\caption{Linear compressibility as a function of temperature for
different system sizes: 216, 512 and 1000 particles. The measuring time
was $3\times 10^6$ md steps in all cases. Other parameters are the same as
in Figure \protect\ref{fig:glass_lin_susc}.}
\label{fig:glass_size_dep}
\end{Fig}
The behavior of the linear generalized compressibility
seen in Figure \ref{fig:glass_lin_susc} is similar to
that seen in measurements of the real part of the frequency dependent
dielectric function $\varepsilon^{\prime}(\omega)$ \cite{Menon95}. In
that case as the frequency decreased, the temperature of the peak in
$\varepsilon^{\prime}(\omega)$ decreased and the drop in
$\varepsilon^{\prime}(\omega)$ below the peak became more abrupt.  By
extrapolating their data to $\omega=0$,  Menon and Nagel \cite{Menon95}
argued that $\varepsilon^{\prime}(\omega=0)$ should diverge at the glass
transition, signaling a second order phase  transition. We have looked
for evidence of this divergence by examining samples of different
sizes to see if the linear generalized
compressibility increased systematically
with system size. As shown in Figure \ref{fig:glass_size_dep} we find
no size dependence and no indication of a diverging linear generalized
compressibility. However, as we mentioned earlier, this does not
preclude the possibility that a thermodynamic phase transition occurs
below $T_{drop}$. It simply means that if there is a growing
correlation length, it is smaller than the size of our system at
$T>T_{drop}$. Another possible reason for the absence of size 
dependence may be that if there is an underlying thermodynamic
phase transition, then its order parameter may not couple to the local
density $\rho(\vec{r})$.

So far we have shown the results of cooling the system. In order
to look for hysteretic behavior we have done runs in which we
heat a system of 512 particles by starting at our lowest temperature
$T=0.1$ with a configuration obtained by cooling
the system. We then increased the temperature in steps of
$\Delta T=0.05$.  As before we equilibrate at each temperature for
$10^4$ time steps and then measure quantities for an additional $10^6$
time steps. Our results are shown in Figure \ref{fig:glass_warm_up}.
Notice the slight hysteresis with the rise in $\chi_{l}$ upon
warming being at a slightly higher temperature than the drop in
$\chi_{l}$ upon cooling. This hysteresis is consistent with 
the kinetic arrest of motion and with the hysteresis found for
the specific heat in Figure \ref{fig:sphtCoolHeat}. 
\begin{Fig}{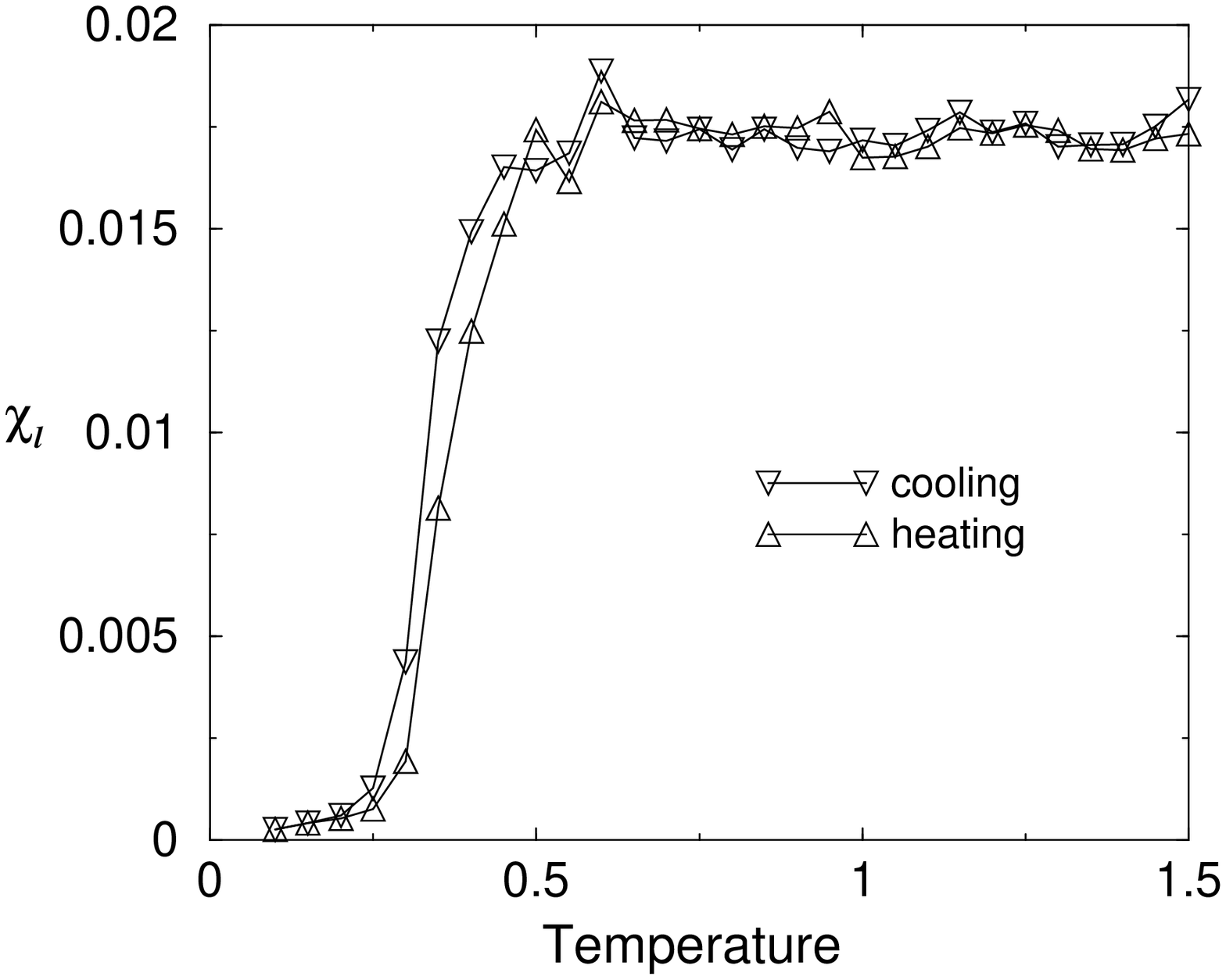}
\caption{Linear generalized
compressibility as a function of temperature for a binary mixture
of 512 particles upon cooling and heating. The measuring time was
$10^6$ md steps in both cases. The data was averaged over 10 runs.
Other parameters are the same as in Figure
\protect\ref{fig:glass_lin_susc}.}
\label{fig:glass_warm_up}
\end{Fig}

\subsubsection{$\chi_{l}$ from Cosine of Positions versus Temperature}
We now consider calculating the linear generalized compressibility
from the cosine transform of the density using eqs. (\ref{eq:susceptibilities})
and (\ref{eq:cosine}). So if we apply a cosine potential along, say the
$\mu=x$ direction, then
\begin{equation}
\rho_{\phi}=\sum_i\cos(k_x x_i)
\end{equation}
where the wave vector
$k_x=2\pi n/L$ with $n=1,2,...$. The wave vectors are compatible with
the periodic boundary conditions of our simulations.
Since the system is isotropic,
we average $\chi_l$ over the $x$, $y$, and $z$ directions. 
The resulting linear
generalized compressibility is qualitatively similar in its temperature
dependence to the linear
compressibility calculated using the absolute values of the
particles' positions (eq. (\ref{eq:abs})).
Figure \ref{fig:coslin_8_1M} shows the linear generalized
compressibility versus temperature for various values of the wave vector.
The data is for a binary mixture of 512 particles with a measuring time of 
$10^6$ md steps and averaged over 10 runs.
\begin{Fig}{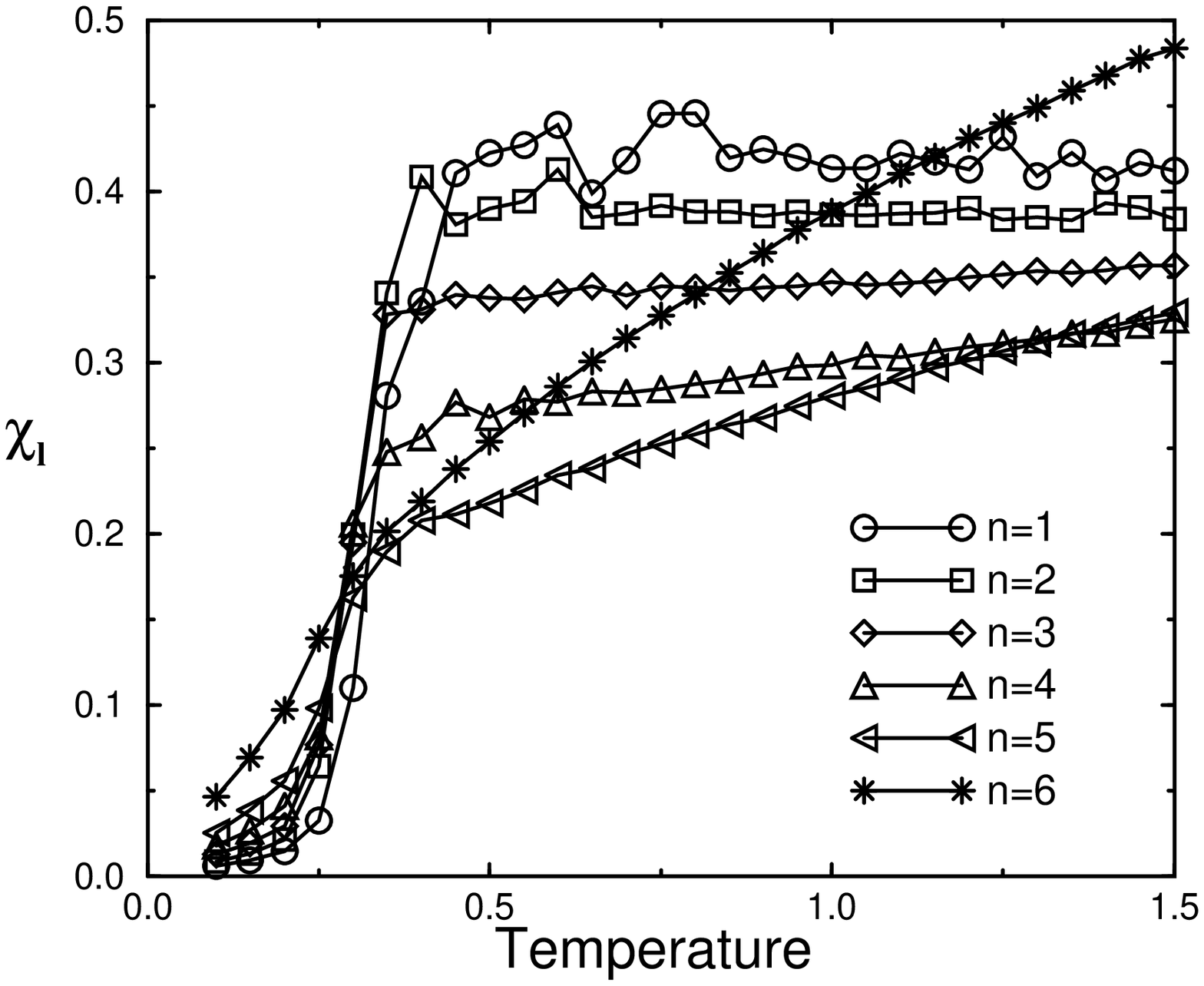}
\caption{Linear generalized
compressibility as a function of temperature for a binary mixture
of 512 particles for different values of the wave vector 
$k=2\pi n/L$. The measuring time was
$10^6$ md steps in all cases. The data was averaged over 10 runs.
The susceptibility was calculated using 
eqs. (\protect\ref{eq:susceptibilities}) and (\protect\ref{eq:cosine}).
Other parameters are the same as in Figure
\protect\ref{fig:glass_lin_susc}.}
\label{fig:coslin_8_1M}
\end{Fig}
Just as for the absolute value case, 
we find that as we increase the measuring time, the drop in the linear
generalized compressibility calculated using cosine
becomes sharper at the glass transition. This is shown in Figure
\ref{fig:coslintime} which shows $\chi_{l}$ as a function of temperature
for measuring times of $2\times 10^5$, $10^6$, and $3\times 10^6$ md
steps with $k=2\pi/L$, i.e., $n=1$.
\begin{Fig}{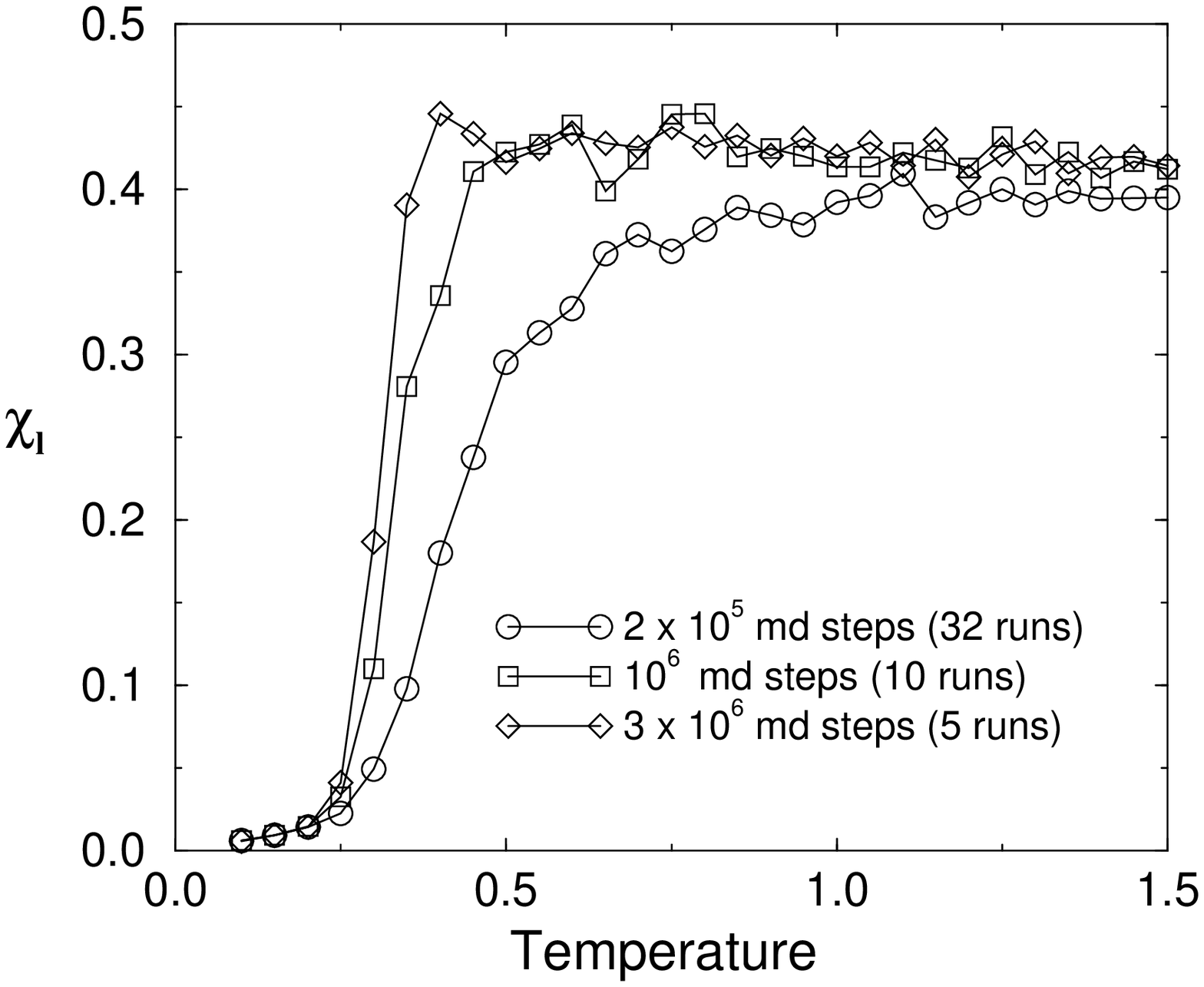}
\caption{Linear generalized
compressibility as a function of temperature for a binary mixture
of 512 particles for different values of the measuring time.
The measuring times are $2\times 10^5$, $10^6$, and $3\times 10^6$
md steps. The data was averaged over the number of runs indicated in the
legend. The susceptibility was calculated using
eqs. (\protect\ref{eq:susceptibilities}) and (\protect\ref{eq:cosine}).
The wave vector $k=2\pi/L$, i.e., $n=1$. 
Other parameters are the same as in Figure \protect\ref{fig:coslin_8_1M}.}
\label{fig:coslintime}
\end{Fig}
Figure \ref{fig:coslinVsk} shows the linear generalized susceptibility
versus the wave vector $k$ in units of $2\pi/L$ for various temperatures.
Note that the dependence is nonmonotonic.
\begin{Fig}{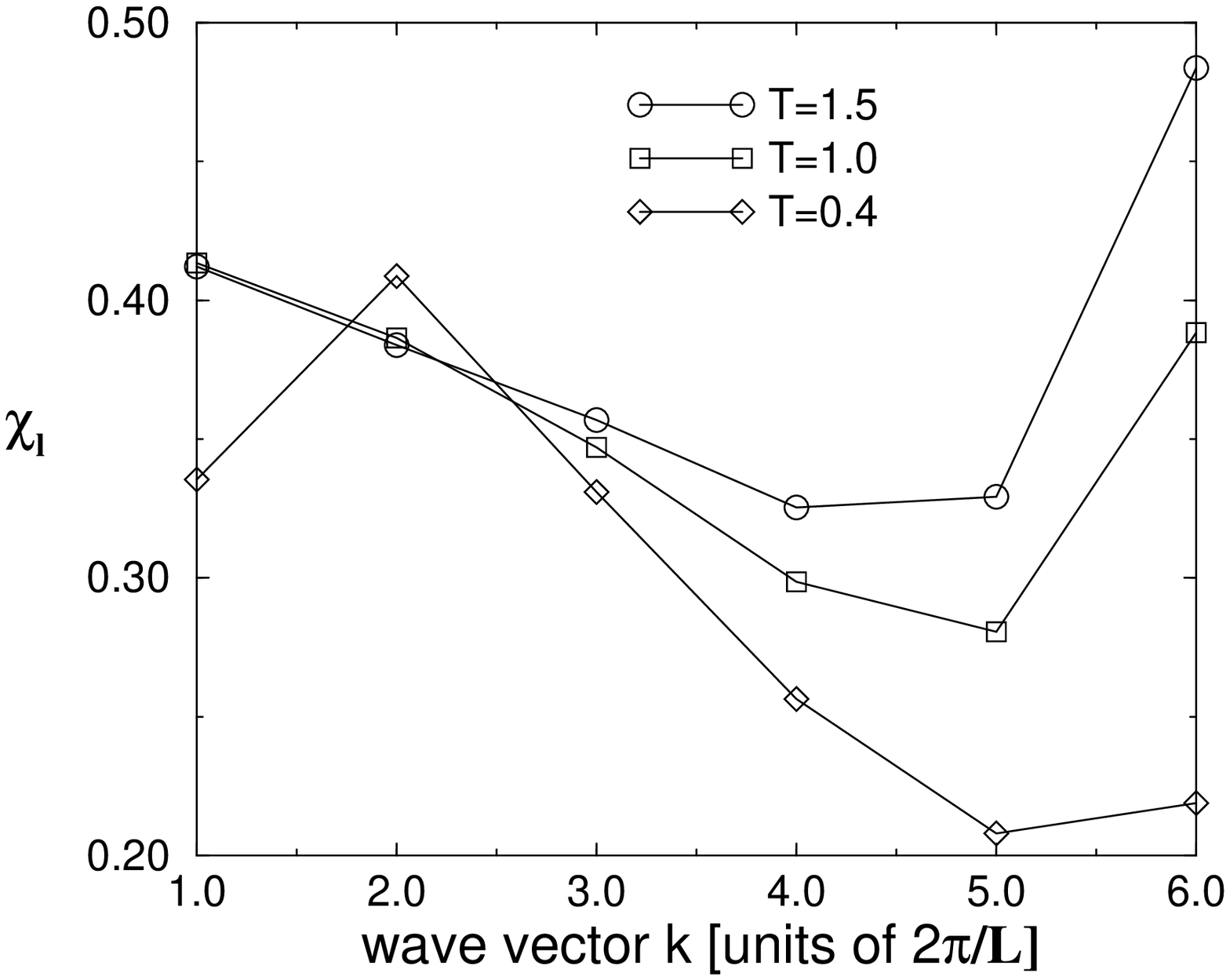}
\caption{Linear generalized
compressibility as a function of wave vector for a binary mixture
of 512 particles for different values of the temperature.
The temperature is measured in md units.
The measuring time was $10^6$ md steps in all cases. The data
was averaged over 10 runs.
The susceptibility was calculated using
eqs. (\protect\ref{eq:susceptibilities}) and (\protect\ref{eq:cosine}).
Other parameters are the same as in Figure
\protect\ref{fig:coslin_8_1M}.}
\label{fig:coslinVsk}
\end{Fig}

\subsubsection{$\chi_{l}$ from Absolute Value of Positions versus Density}
In Figure \ref{fig:abslindens} we plot the generalized linear
compressibility versus density calculated from the absolute
value of the positions of the particles.
In Figure \ref{fig:coslindens} we plot the generalized linear
compressibility versus density calculated from the cosine
of the particles' positions. In both cases we see that $\chi_l$
drops with increasing density. The drop becomes more abrupt
as the measuring time increases. This drop is similar to
what we see when we cool the system at fixed density.
The density ($\rho\approx 0.8$, $\Gamma\approx 1.44$)
at which the drop occurs agrees with the density
at which there is a peak in the specific heat as shown
in Fig. \ref{fig:sphtdens}. 

\begin{Fig}{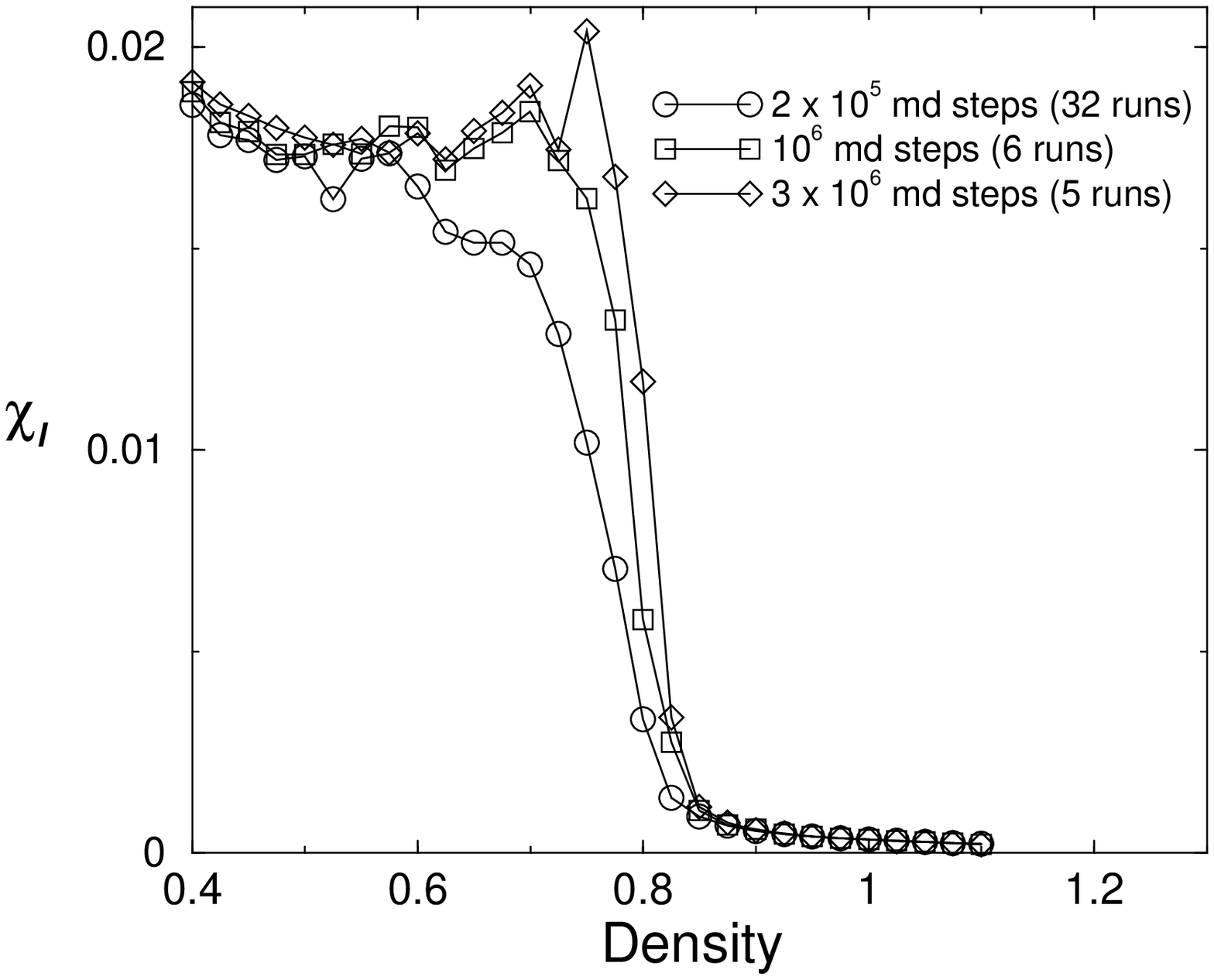}
\caption{Generalized linear compressibility
as a function of density for a binary
mixture of 512 particles at $T=1$. The measurement times are
$2\times 10^{5}$, $10^6$ and $3\times 10^{6}$ md steps.
$\sigma_B/\sigma_A=1.4$. Data was averaged over the number of 
runs indicated in the legend. $\chi_{l}$ is
calculated using the absolute value of the particles' positions.}
\label{fig:abslindens}
\end{Fig}

\begin{Fig}{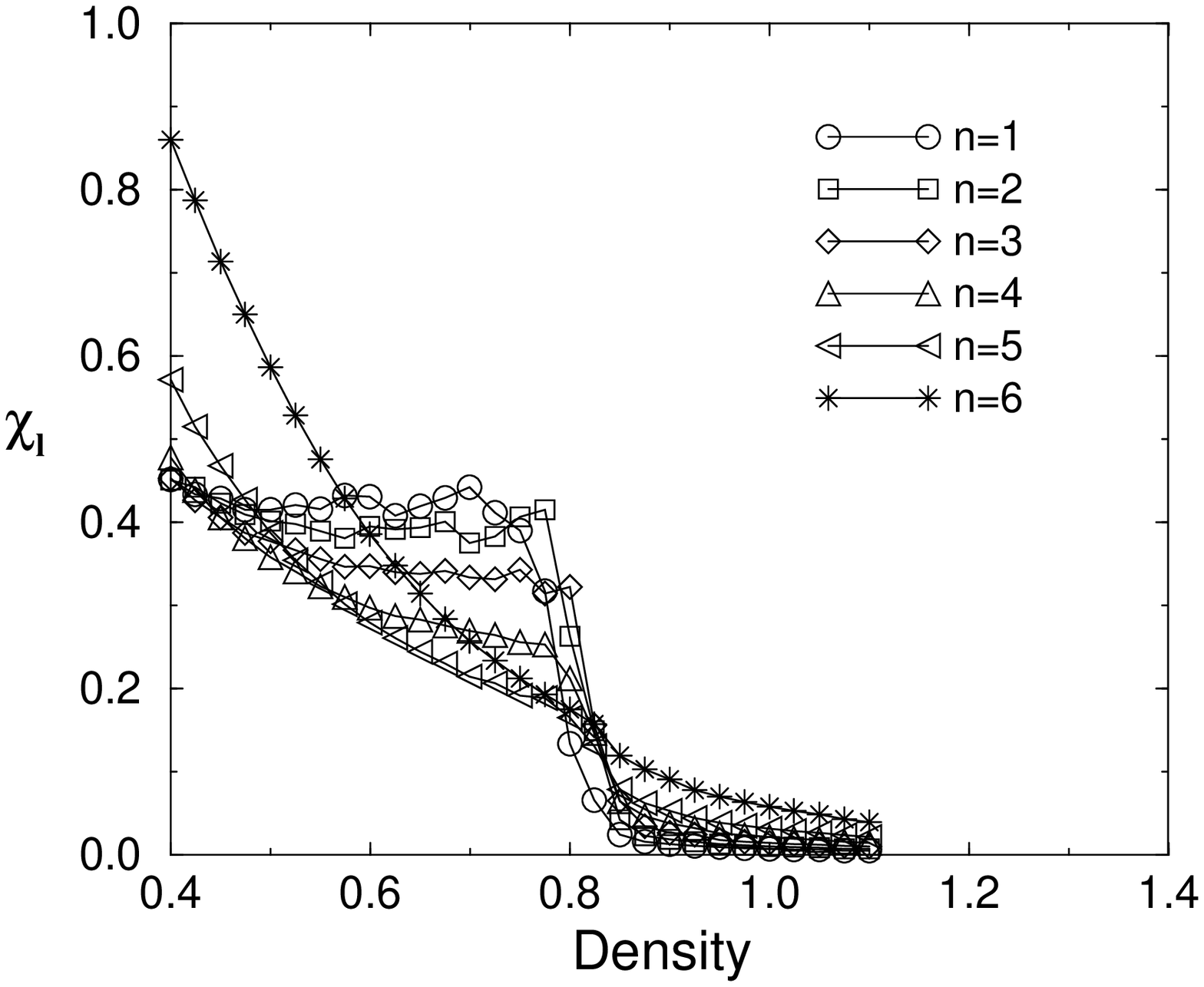}
\caption{Generalized linear compressibility
as a function of density for a binary
mixture of 512 particles at $T=1$. The different values of
$n$ correspond to different values of the wave vector
$k=2\pi n/L$. The measurement time is
$10^6$ md steps in all cases. 
$\sigma_B/\sigma_A=1.4$. Data was averaged over 6
runs. The susceptibility was calculated from the cosine of the particles'
positions using
eqs. (\protect\ref{eq:susceptibilities}) and (\protect\ref{eq:cosine}).
Other parameters are the same as in Figure
\protect\ref{fig:coslin_8_1M}.}
\label{fig:coslindens}
\end{Fig}

\subsubsection{$\chi_{l}$ from Absolute Value of Positions versus 
Temperature in a Slab Geometry}
So far we have considered systems with a fixed number of particles,
but as we mentioned earlier in this section, we can generalize our
results to the grand canonical ensemble where the number $N$ of 
particles can vary. We have examined the generalized linear
compressibility $\chi_l$ calculated from the absolute value of the
particle positions using equations (\ref{eq:susceptibilities}) 
and (\ref{eq:abs}) for a slab of our system. In other words we have
divided a system of $8^3$ particles into 8 slabs of equal thickness
perpendicular to the $x$ axis. The number of particles in any given 
slab is not fixed. However, in eq. (\ref{eq:susceptibilities})
we set the average number $N$ of particles in each slab equal to the
total number of particles in the system divided by the number of layers,
i.e., $N=8^3/8=64$. Such a slab geometry mimicks experiments
on colloidal suspensions of binary mixtures in which the focal 
plane of the camera can essentially see only one monolayer of 
polystyrene balls \cite{comment:cherry}. 
In figure \ref{fig:slab} we show the generalized
linear compressibility for a slab for two different measuring times. Again
we see that the drop is sharper as the measurement time becomes longer.
Thus allowing for fluctuations in the number of particles does not
change the qualitative behavior of $\chi_l$ at the glass transition.
Comparing Figures \ref{fig:glass_lin_susc} and \ref{fig:slab}, we 
see that the temperature of the drop for the slab and the
bulk agree. We also notice that the drop is sharper for the bulk where
presumably the greater number of particles results in
better statistical averaging.
\begin{Fig}{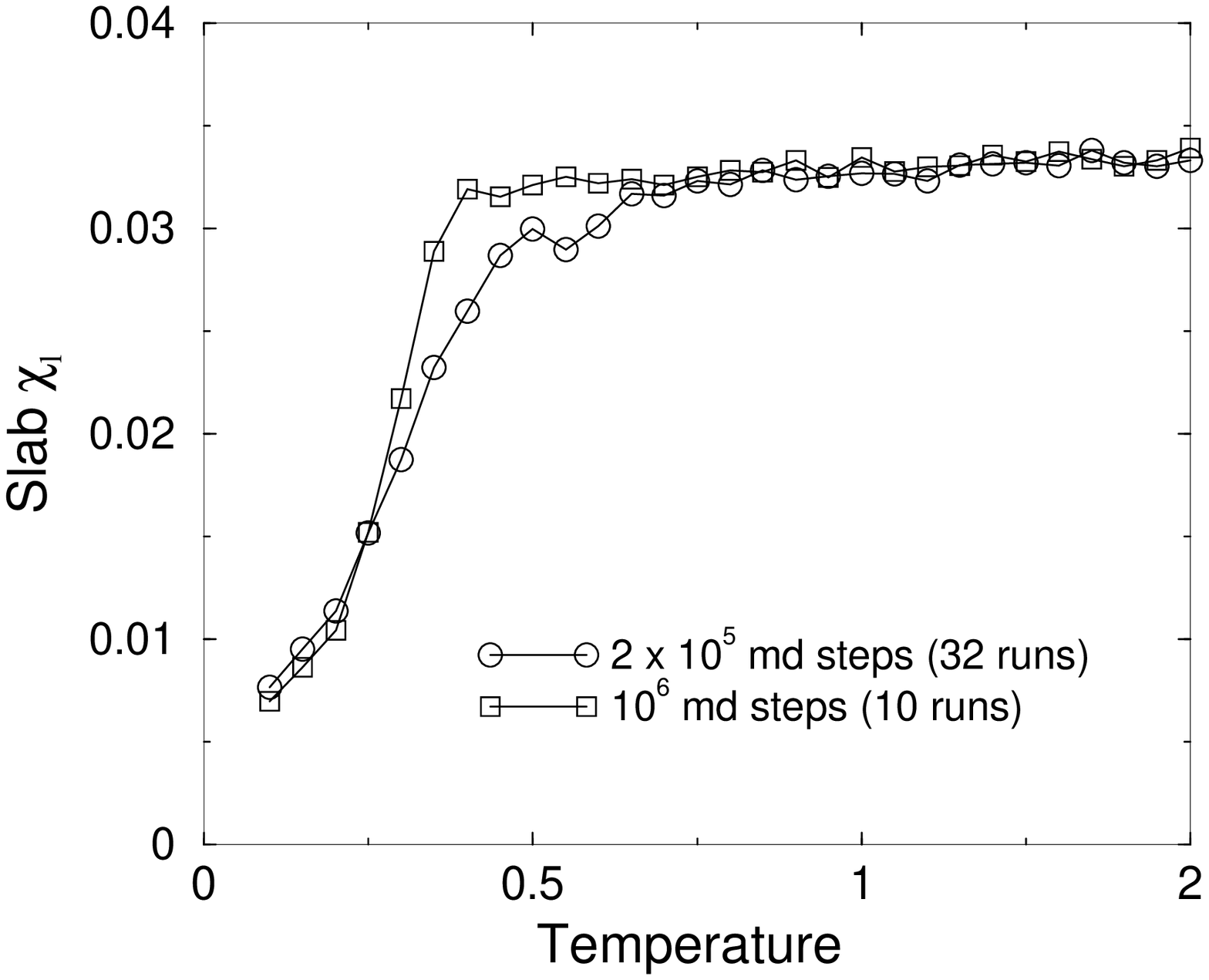}
\caption{Linear generalized
compressibility as a function of temperature for a monolayer
slab in a binary mixture
of 512 particles for different values of the measuring time.
The measuring times are $2\times 10^5$ and $10^6$
md steps. The data was averaged over the number of runs indicated in the
legend. The susceptibility was calculated using
eqs. (\protect\ref{eq:susceptibilities}) and (\protect\ref{eq:abs}).
Other parameters are the same as in Figure \protect\ref{fig:glass_lin_susc}.}
\label{fig:slab}
\end{Fig}

\subsubsection{$\chi_{l}$ from Absolute Value of Positions versus Density in a
Slab Geometry}
Since experiments on colloidal suspensions usually vary the density
rather than the temperature, we have done simulations where we set
the temperature $T=1$ and vary the density. Again we divide
our system of $N=8^3=512$ particles into 8 slabs and measure
$\chi_l$ in one of those slabs. The results are shown in
Figure \ref{fig:slabdens}.
\begin{Fig}{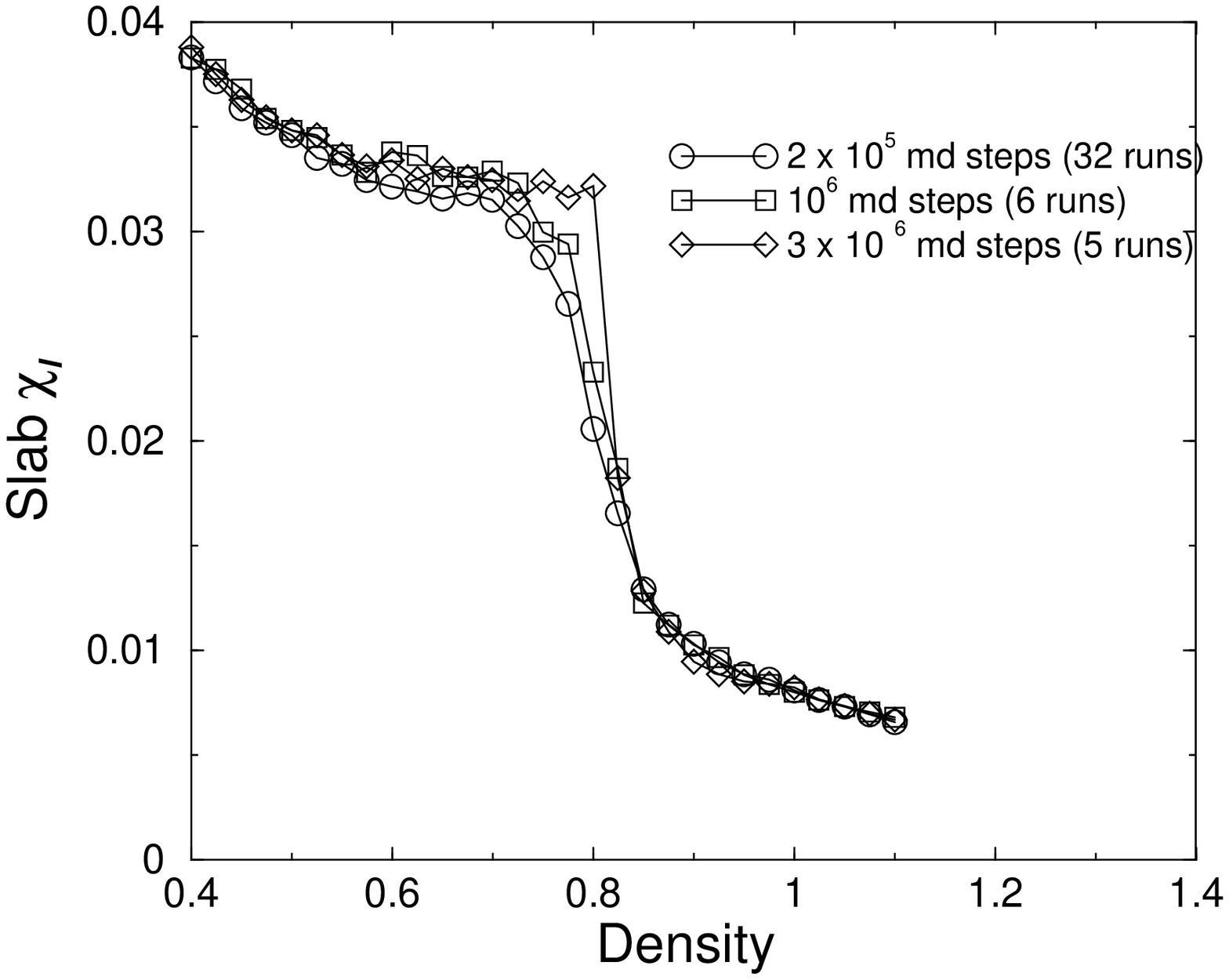}
\caption{Linear generalized compressibility as a function of density
for a monolayer slab in a binary mixture
of 512 particles for different values of the measuring time.
The measuring times are $2\times 10^5$, $10^6$ and $3\times 10^6$
md steps. The data was averaged over the number of runs indicated in the
legend. The susceptibility was calculated using
eqs. (\protect\ref{eq:susceptibilities}) and (\protect\ref{eq:abs}).
Other parameters are the same as in Figure \protect\ref{fig:glass_lin_susc}.}
\label{fig:slabdens}
\end{Fig}
As one can see from the figure, $\chi_l$ drops as the density is increased
and the drop becomes more abrupt as the measuring time lengthens.
Comparing Figures \ref{fig:abslindens}, \ref{fig:coslindens} 
and \ref{fig:slabdens}, we see that the drop occurs at the same density
($\rho\approx 0.8$) as in the bulk. 

\subsection{Nonlinear Generalized Compressibility}
We now turn to the case of the nonlinear generalized compressibility
$\chi_{nl}$ given by eq. (\ref{eq:susceptibilities}). We are
motivated by the case of spin glasses where 
the nonlinear magnetic compressibility diverges at the spin glass
transition while the linear compressibility only has a cusp 
\cite{Bhatt88,Levy86}.
There have been a few studies of nonlinear response functions in structural
glasses \cite{Dasgupta91,Wu91}, but these have not found any
divergences.  Our results are consistent with this conclusion. In
particular we find that the nonlinear generalized compressibility is
zero above and below the glass transition temperature, though it does show a
glitch at the glass transition temperature. There is no systematic increase
with system size, indicating the absence of a divergence at temperatures
above the glass transition. This does not rule out a divergence
below the glass transition temperature where our system has fallen 
out of equilibrium. It also does not rule out a thermodynamic transition
that does not couple to the local density. Because $\chi_{nl}$ is sensitive to
the tails of the distribution of $\rho_{\phi}$, one must be careful to
obtain a good ensemble average in the liquid above the glass
transition temperature. We have done this by doing 16 or 32 runs, each
involving 200,000 time steps, with different initial conditions,
stringing them together as though they were one long run and then
taking the appropriate averages. In some sense this approach
mixes molecular dynamics and Monte Carlo; the simulation 
follows the equations of motion for a given amount of time 
and then ``jumps'' to another configuration which again evolves according
to the molecular dynamics equations until the next jump. We call this approach
``global averaging.'' It produces a better ensemble
average of $<\rho_{\phi}^2>^2$ which enters into $\chi_{nl}$ in
eq. (\ref{eq:CUMU4}). The resulting $\chi_{nl}$ is shown in 
Figure \ref{fig:nonlinGlobal}
which was calculated from the absolute values of the particles' positions
using eqs. (\ref{eq:susceptibilities}) and (\ref{eq:abs}). 
\begin{Fig}{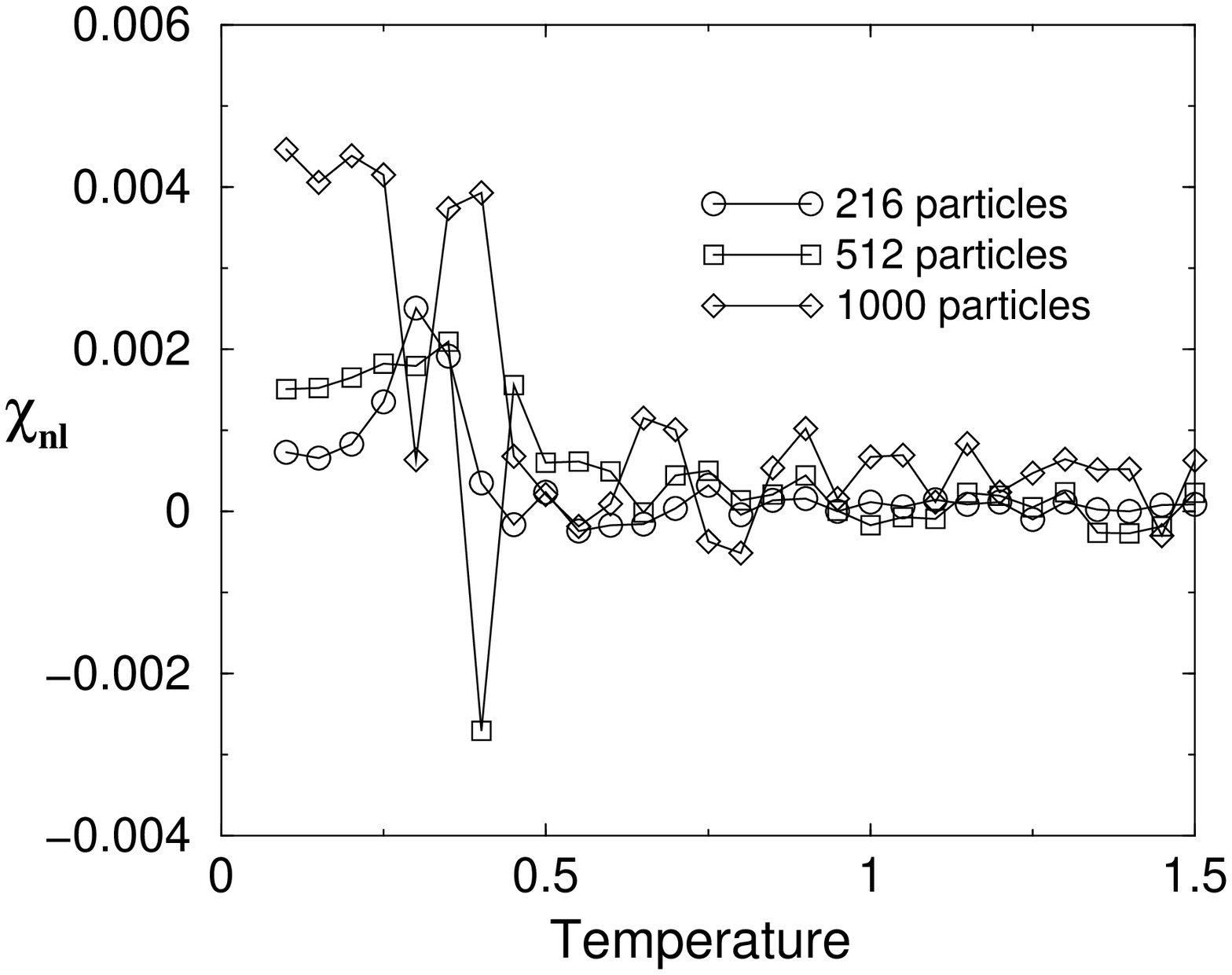}
\caption{Nonlinear generalized compressibility as a function of 
temperature for binary mixtures. The data for 216 particles is from
$3.2\times 10^6$ md steps 
obtained by stringing together 16 runs, each of which involved $2\times 10^5$
md steps. The data for 512 and 1000 particles is from $6.4\times 10^6$
md steps obtained by stringing together 32 runs of
$2\times 10^5$ md steps. $\rho_o = 0.6$ and $\sigma_B/\sigma_A=1.4$.}
\label{fig:nonlinGlobal}
\end{Fig}

$\chi_{nl}$ also took longer to equilibrate
than $\chi_{l}$. By plotting $\chi_{nl}$ versus run time, we
found that one had to run at least $10^6$ time steps at $T=1$ before
$\chi_{nl}$ appeared to saturate (see Figure \ref{fig:nonlinTime}).
\begin{Fig}{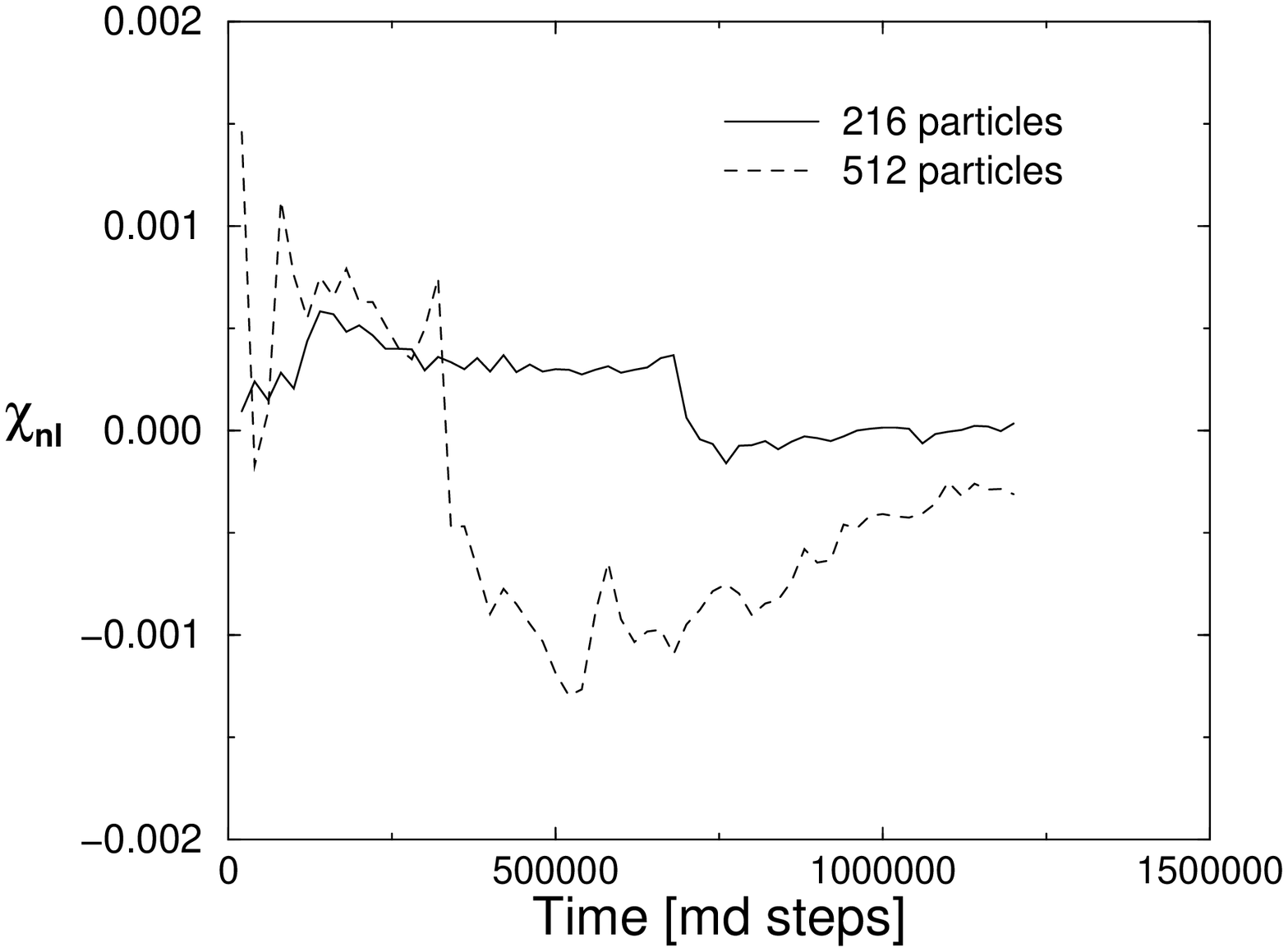}
\caption{Nonlinear generalized compressibility as a function of
time for binary mixtures of 216 and 512 particles at $T=1$. The data shown
for each system size is for a single run and a running average is kept. 
$\chi_{nl}$ is calculated using eqs. (\protect\ref{eq:susceptibilities}) 
and (\protect\ref{eq:abs}). $\rho_o = 0.6$ and $\sigma_B/\sigma_A=1.4$.}
\label{fig:nonlinTime}
\end{Fig}

\section{Ordinary Isothermal Compressibility}
The ordinary isothermal compressibility $\kappa_{T}$ can be related to
the fluctuations in the number, volume or density of the system. In most 
of our calculations we fix the volume, number of particles and density, so
that there are no such fluctuations and the system has $\kappa_{T}=0$.
However in the grand canonical ensemble $\kappa_{T}$ is given by
\begin{equation}
\kappa_{T}=\frac{1}{\rho_{o}k_BT}
\frac{\langle N^2\rangle-\langle N\rangle^2}{\langle N \rangle}
\end{equation}
where $\rho_o=<N>/V$. To relate $\kappa_{T}$ to the linear generalized
compressibility $\chi_l$, we choose a uniform potential $\phi(\vec{r})=1$. Then
$\rho_{\phi}= \int d^3r \rho(\vec{r})=N$ and
\begin{equation}
\kappa_{T}=\frac{1}{6\rho_{o}k_BT}\chi_{l}
\end{equation}

In principle one can also obtain $\kappa_{T}$ from the $k\rightarrow 0$
limit of the static structure factor 
$S({\vec{k}})=(1/N)\langle\rho_{{\vec{k}}}\rho_{-{\vec{k}}}\rangle$
where $\rho_{\vec{k}}=\sum_{i=1}^{N}\exp(-i\vec{k}\cdot\vec{r}_{i})$
is the Fourier transform of the local density $\rho(\vec{r})$. The 
limit of $S({\vec{k}})$ for $k\rightarrow 0$ in an isotropic and
homogeneous system is \cite{Hansen}
\begin{eqnarray}
S(0) &=& 1 + \rho_o\int(g(r)-1)d^{3}r \nonumber\\
&=& \rho_o k_B T\kappa_{T}
\label{eq:SkT}
\end{eqnarray}
where $g(r)$ is the radial distribution function.
Note that in a system with fixed volume and particle number, the
normalization of $g(r)$ leads to $S(k\rightarrow 0)=0$. This is 
consistent with the fact that such a system 
has $\kappa_{T}=0$. Eq. (\ref{eq:SkT}) yields a nonzero value for $\kappa_{T}$
in a system which has fluctuations in volume,
particle number or density. Even in such a compressible system
taking the $k\rightarrow 0$ limit of $S(k)$ can suffer from finite size
effects \cite{Bernu85}
because the farthest apart that any two particles can be along
any given coordinate axis is $L/2$ when there are periodic boundary
conditions. So at wave vectors $k$ with components smaller
than 4$\pi/L$, $S(k)$ can have spurious results. (For example
in our simulations we found that this manifests itself as a slight
upturn in $S(k)$ at small $k$.) It is possible to
extrapolate to distances larger than $L/2$ using various approaches
\cite{Bernu85}. We chose not to use this approach to calculate
$\kappa_T$ since we work in a system with fixed $N$ and $V$.
We should note however that simulations \cite{Horbach99} working in 
the $NVT$ ensemble with fixed $N$, $V$, and $T$, have successfully used 
eq. (\ref{eq:SkT}) to find the isothermal compressibility. We can
resolve this with the fact that $S(k\rightarrow 0)=0$ in the $NVT$
ensemble by noting that for values of $k>4\pi/L$, $S(k)$ should
give the same value in the $NVT$ ensemble as in the grand canonical
ensemble. So if $L$ is large enough, fitting $S(k)$ to the small
$k$ form $S(k)=S(k\rightarrow 0)+Ak^2$, where $A$ is a constant,
should yield the correct value of $\kappa_T$ as long as $k>4\pi/L$.

Rather than using eq. (\ref{eq:SkT}),
we calculated $\kappa_T$ by monitoring a small subvolume
inside of our system and keeping track of the fluctuations in the number
of particles in the subvolume. Let us
define a dimensionless ordinary isothermal compressibility $K_{T}$
by 
\begin{equation}
\kappa_{T}=\frac{1}{\rho k_BT}K_T
\end{equation}
where
\begin{equation}
K_T=\frac{\langle N^2\rangle - \langle N\rangle^2}{\langle N\rangle}
\label{eq:KT}
\end{equation}
In order to calculate $K_T$,
we have monitored a subvolume that had on average 25\% of the
total number of particles. Essentially we drew an imaginary
boundary in the middle of our system that enclosed
25\% of the total volume and counted the number of particles in
this subvolume as a function of time. By monitoring the fluctuations in the
number of particles in this subvolume, we could calculate
$K_T$. The results for a subvolume which had on average 
128 particles out of a total of 512 particles are shown in
Figure \ref{fig:realcompress} where $K_T$ is plotted versus temperature.
We see that it has the same basic shape as the linear generalized
compressibililty with a drop at the same temperature as $\chi_l$.
As with $\chi_l$, the drop becomes sharper with increasing measuring
time.
\begin{Fig}{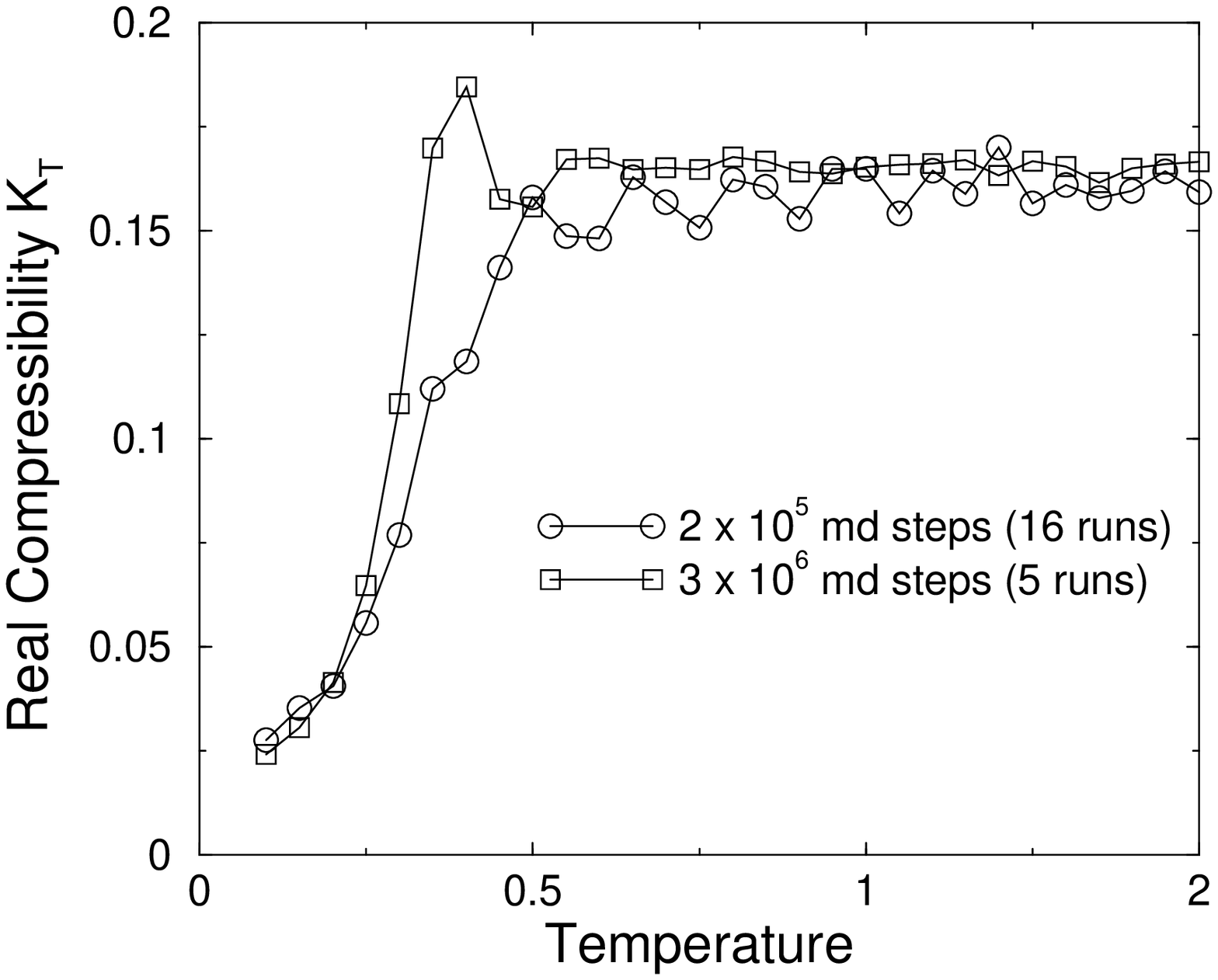}
\caption{Dimensionless ordinary isothermal compressibility $K_T$
as a function of temperature for a subset of a system 
of 512 particles for different values of the measuring time.
On average the subset had 128 particles in it.
The measuring times are $2\times 10^5$ and $3\times 10^6$
md steps. The data was averaged over the number of runs indicated in the
legend. The dimensionless compressibility was calculated using
eq. (\protect\ref{eq:KT}). 
Other parameters are the same as in Figure \protect\ref{fig:glass_lin_susc}.}
\label{fig:realcompress}
\end{Fig}

While $K_T$ shows behavior similar to the linear generalized
compressibility $\chi_{l}$ as a function of temperature for a given size,
it is unlike $\chi_{l}$ in that it suffers from finite size effects.
We have demonstrated this by making measurements on systems with
a total of 64, 216, 512, and 1000 particles. The measurements were
made by counting the number of particles in a subvolume that was
25\% of the total volume. Such a small subvolume has a large
surface to volume ratio which produces large finite
size effects. To understand this, we note that
in such a small subvolume, a significant
number of the particles are very close to the boundary of the
subvolume. Fluctuations in the positions of these particles moves
them in and out of the subvolume, producing large fluctuations
in the number of particles in the subvolume. The smaller the
system, the bigger this effect is. This produces large finite
size effects in $K_T$ even at high temperatures
where the system easily equilibrates. This can be seen
in Figure \ref{fig:realcompressize}. One can see that 
$K_T$ decreases with increasing system size
at high temperatures above the drop in the compressibility.
One of the advantages of the linear generalized compressibility
is the absence of these finite size effects. In fact the linear
generalized compressibility shows no size dependence at temperatures
above the observed glass transition (see Figure \ref{fig:glass_size_dep}).
\begin{Fig}{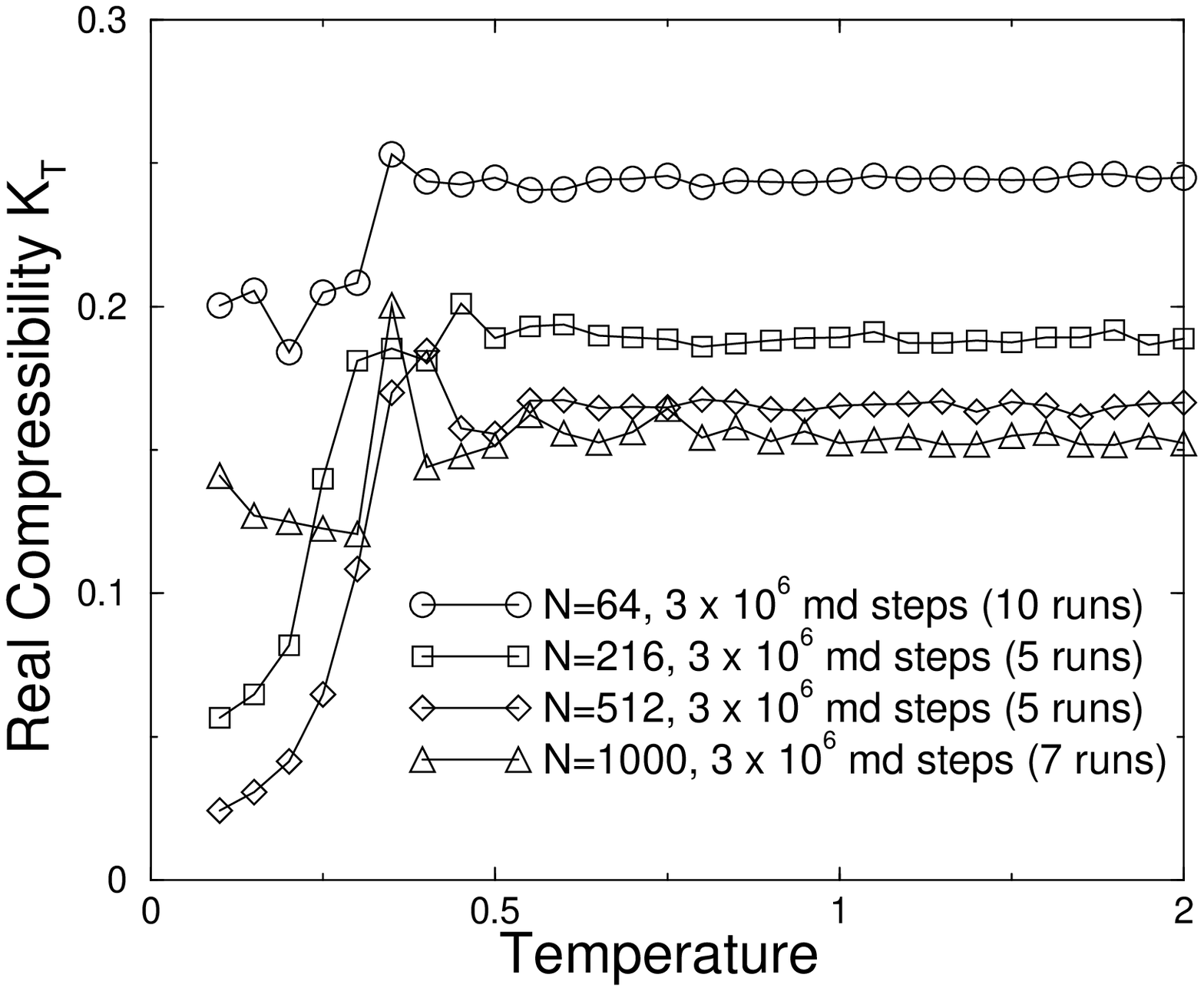}
\caption{Dimensionless ordinary isothermal compressibility $K_T$
as a function of temperature for systems with a total of
$N$ = 64, 216, 512 and 1000 particles. The isothermal compressibility
was obtained by monitoring a
subvolume that had on average 25\% of the particles in it.
Notice the size dependence at high temperatures.
The measuring time is $3\times 10^6$
md steps. The data was averaged over the number of runs indicated in the
legend. The dimensionless compressibility was calculated using
eq. (\protect\ref{eq:KT}). 
Other parameters are the same as in Figure \protect\ref{fig:glass_lin_susc}.}
\label{fig:realcompressize}
\end{Fig}

\section{Diffusion Constant}
The diffusion of the particles reflects the kinetics of the system
and becomes very small below the 
glass transition temperature.
We calculate the diffusion constant $D$ using the Einstein relation
\begin{equation}
D=\lim_{t\rightarrow\infty}{1\over 6Nt}
\langle \sum_{i=1}^{N}({\bf r}_i(t)-{\bf r}_i(0))^2\rangle
\label{eq:diffusion}
\end{equation}
where ${\bf r}_i$ are true displacements of the $i$th particle.
Since we are using periodic boundary conditions, if the particle has 
crossed the box several times, then this must be included in 
${\bf r}_i$. 

\subsection{Diffusion Constant Versus Temperature}
As the system is cooled through the glass transition, the diffusion constant
calculated using equation (\ref{eq:diffusion}) becomes very small.
This is shown in Figure \ref{fig:diffLog} where the diffusion constant
for 512 particles is plotted on a logarithmic scale. 
The diffusion constant varies smoothly over the entire temperature range.
The curves corresponding to different cooling rates begin to separate
as the system falls out of equilibrium at the glass transition temperature
where the specific heat peaks and where the linear generalized
compressibility drops abruptly.
Figure \ref{fig:diffLog} also shows the diffusion constant for binary
mixtures of several different sizes. Notice that there is no apparent
size effect.
\begin{Fig}{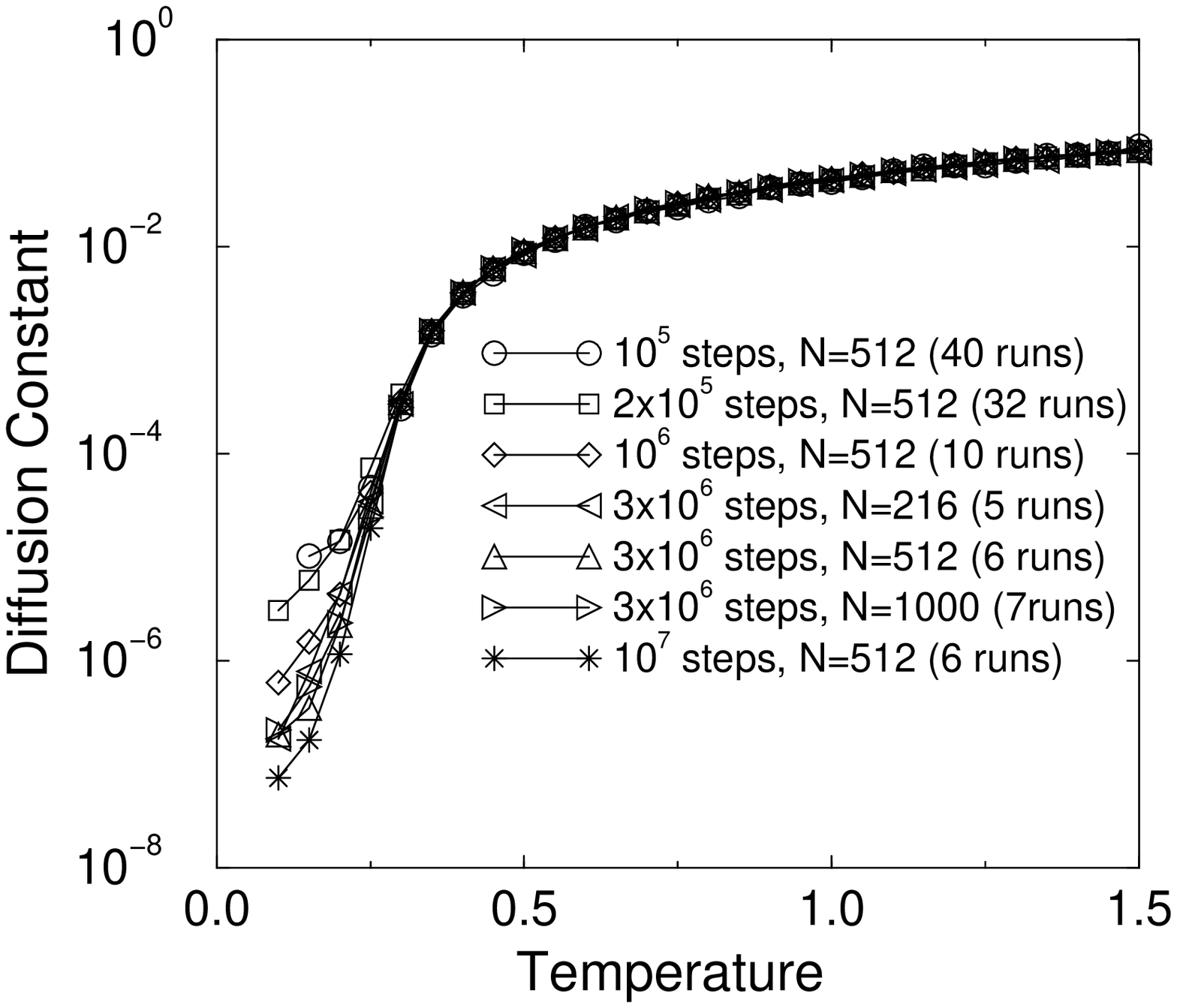}
\caption{Diffusion constant as a function of temperature for a binary
mixture of particles plotted on a logarithmic scale. 
Measurement times for 512 particles are $10^5$, $2\times 10^5$,
$10^6$, $3\times 10^6$, and $10^7$ md steps. Measurement time
for 216, 512, and 1000 particles is $3\times 10^6$ md steps.
Number of runs averaged over is given in the legend. 
$\rho_o = 0.6$ and $\sigma_B/\sigma_A=1.4$.}
\label{fig:diffLog}
\end{Fig}

\subsection{Diffusion Constant Versus Density}
In Figure \ref{fig:diffdens} we show the diffusion constant as a function
of density. We see that the diffusion decreases smoothly as the
density $\rho$ increases. The curves corresponding to different
measurement times begin to separate at the glass transition density
where the specific heat peaks and where the linear generalized
compressibility drops abruptly.
\begin{Fig}{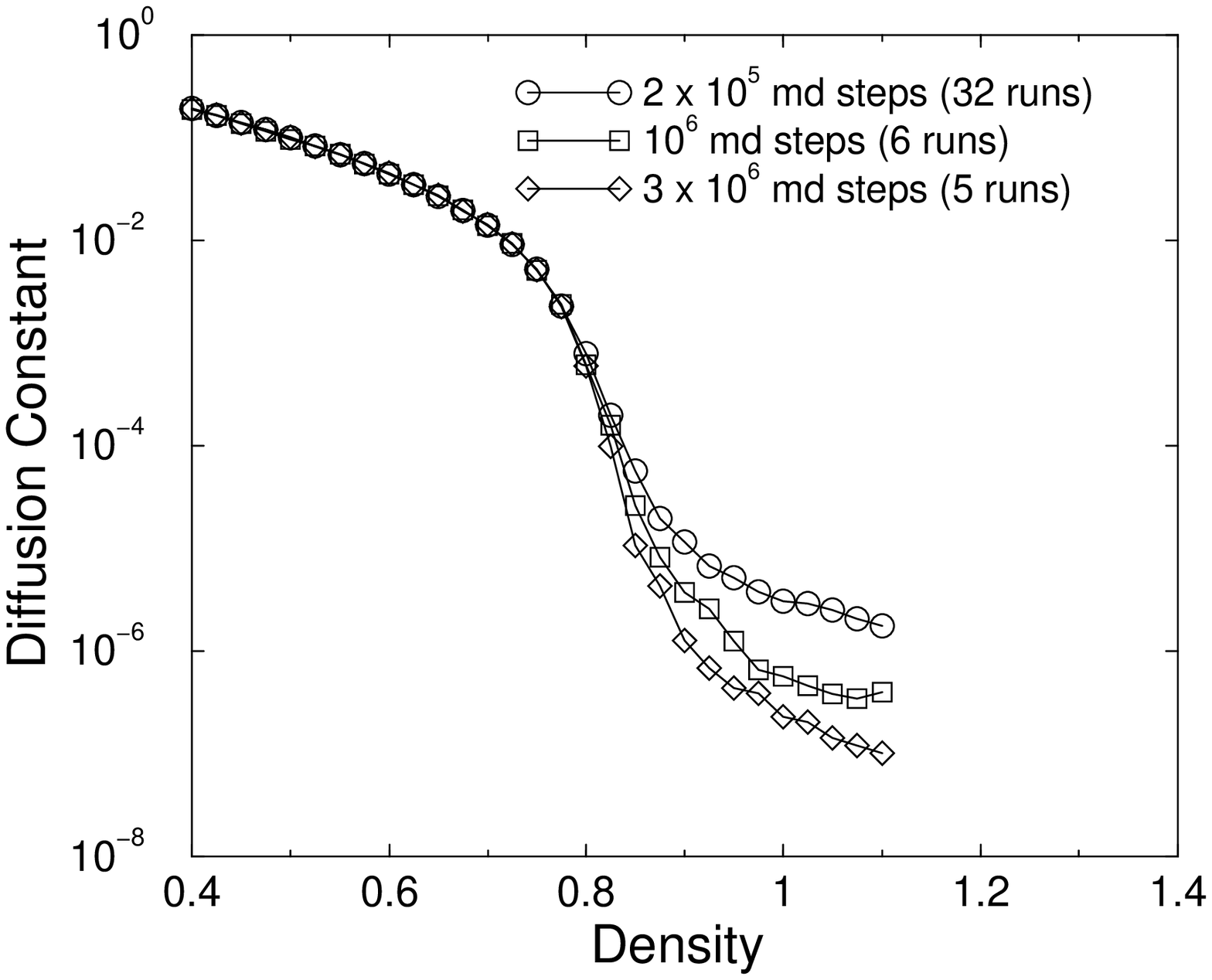}
\caption{Logarithmic plot of the diffusion constant as a function of density 
for a binary mixture of 512 particles at $T=1$. The measurement times are
$2\times 10^{5}$, $10^6$ and $3\times 10^{6}$ md steps. 
$\sigma_B/\sigma_A=1.4$. Data
was averaged over the number of runs indicated in the legend.}
\label{fig:diffdens}
\end{Fig}

\section{Relation between $\chi_{l}$ and $S(k)$}
When the system has translational invariance,
we can relate the linear generalized compressibility $\chi_{l}$ to 
the static structure factor $S(k)$ which is measured in experiments such
as neutron scattering. $S(k)$ is also used as an
input for mode coupling theories 
\cite{Gotze92} and it is generally assumed that $S(k)$ does not show any 
essential variations near the glass transition as the temperature or density is 
varied. As we shall see, our calculation agrees with this.

By definition, the static structure factor
$S({\vec{k}})=(1/N)\langle\rho_{{\vec{k}}}\rho_{-{\vec{k}}}\rangle$.
To relate $S(k)$ to $\chi_{l}$, we first note that $S(k)$ is the
Fourier transform of the static density--density autocorrelation
function $G(\vec{r})$, i.e., 
$S(\vec{k})=\int\exp(-i\vec{k}\cdot\vec{r})G(\vec{r}) d^3r$, 
where $G(\vec{r})=(1/N)\int\langle
\rho(\vec{r}^{\;\prime}+\vec{r})\rho(\vec{r}^{\;\prime})\rangle 
d^3r^{\prime}$.
In the supercooled liquid above the glass transition, the system
has translational invariance, and we can write
\begin{equation}
\langle\rho(\vec{r})\rho(\vec{r}^{\;\prime})\rangle=\frac{1}{V}
g\left(\vec{r}-\vec{r}^{\;\prime}\right)
\end{equation}
where $V$ is volume and $g$ is a function of the difference
$\left(\vec{r}-\vec{r}^{\;\prime}\right)$. In this case 
$G(\vec{r})=g(\vec{r})/N$.
In $\chi_{l}$ we meet 
\begin{equation}
\langle\rho_{\phi}^{2}\rangle=\int d^3r d^3r^{\prime}\phi(\vec{r})
\phi(\vec{r}^{\;\prime})\langle\rho(\vec{r})\rho(\vec{r}^{\;\prime})\rangle
\end{equation}
If there is translational invariance,
\begin{eqnarray}
\langle\rho_{\phi}^{2}\rangle &=&\frac{1}{V}\int d^3r d^3r^{\prime}
\phi(\vec{r})\phi(\vec{r}^{\;\prime})g\left(\vec{r}-\vec{r}^{\;\prime}\right)
\nonumber\\
&=&\frac{1}{V}\int\frac{d^3 k}{(2\pi)^3}\phi(\vec{k})\phi(-\vec{k})
g(-\vec{k})\nonumber\\
&=&\rho_{o}\int\frac{d^3k}{(2\pi)^3}|\phi(\vec{k})|^2 S(\vec{k})
\label{eq:rhoSk}\\
\nonumber
\end{eqnarray}
where $\phi(\vec{k})$ is the Fourier transform of $\phi(\vec{r})$.
Converting the integral $V\int d^3k/(2\pi)^3$ to a sum $\sum_{\vec{k}}$
and using eq. (\ref{eq:susceptibilities}), we obtain
\begin{equation}
\chi_{l}=\frac{6}{N}\left[\frac{N}{V^2}\sum_{\vec{k}}\left(
|\phi(\vec{k})|^2 S(\vec{k})\right)-\langle\rho_{\phi}\rangle^2\right]
\label{eq:chiSk}
\end{equation}

As an example, let us choose
$\phi(\vec{r})=\cos(k_x x)$ with the proviso that $k_x\neq 0$.
(When $\vec{k}=0$, $\chi_{l}=0$ since the potential is uniform
and there are no fluctuations allowed with fixed $N$ and $V$.)
With this choice of $\phi(\vec{r})$,
$\rho_{\phi}=\sum_{i}\cos(k_x x_i)$. 
Then one can show by explicitly calculating $\phi(\vec{k})$ and by using
eqs. (\ref{eq:rhoSk}) and (\ref{eq:chiSk}) that
\begin{equation}
\chi_{l}(\vec{k})=3 S(\vec{k})-
\frac{6}{N}[{\rm Re}\langle\rho_{\vec{k}} \rangle]^2
\label{eq:chi}
\end{equation}
where translational invariance allows us to write
\begin{eqnarray}
S(\vec{k})&=&(2/N)\langle\rho_{\phi}^2\rangle \nonumber \\
&=&(2/N)\sum_{ij}
\left[\langle\cos(k_x x_i)\cos(k_x x_j)\rangle
\right]
\label{eq:Sk}
\end{eqnarray}
Note that the value of $k$ that we use to probe the glass transition is 
typically of order $k_{L}\equiv 2\pi/L$ which is much smaller than the 
value of $k_{\rm peak}$ at which the first peak in $S(k)$ appears. 
$k_{\rm peak}\sim 2\pi/\sigma_{A}\gg k_{L}$ where $\sigma_{A}$ is the 
diameter of type A spheres in the binary mixture. 

We have numerically calculated $[S(k)]_{\rm run}$ for our binary mixture using
equation (\ref{eq:Sk}) with $\phi(\vec{r})=\cos(k_{\mu}r_{\mu})$. 
(No sum over repeated indices. $\mu=x$, $y$, or $z$.)
$[...]_{\rm run}$ is an
average over all the runs and over $\vec{k}$ being parallel to $x$, $y$, and
$z$.
The result is shown in Figure \ref{fig:Sk}
and one can see that $[S(k)]_{\rm run}$ does not vary much through 
the glass transition which is consistent with what is assumed in
mode coupling theory. The figure also shows 
$[{\rm Re}<\rho_{k}>]_{\rm run}$ where $<...>$ is a thermal average 
over a single run. If $[S(k)]_{\rm run}$ 
and $[{\rm Re}<\rho_{k}>]_{\rm run}$ do not vary much
through the glass transition, how can the difference between 
the two terms in (\ref{eq:chi}) decrease and produce a drop in $\chi_l$? 
To answer this, note that there are two inequivalent
ways in which one can calculate
$\chi_l$. So far we have calculated $\chi_l$
for each run and then averaged over the different runs. This
approach is what we used in 
Figs. \ref{fig:glass_lin_susc}--\ref{fig:glass_warm_up} 
and results in a sharp drop in the linear generalized
compressibility at the glass transition. Let us call this
a run--by--run average for which we can write: 
\begin{equation}
\chi_{l}=\frac{6}{N}\left[\langle\rho_{\phi}^2\rangle-
\langle\rho_{\phi}\rangle^2\right]_{\rm run}
\label{eq:runbyrunavg}
\end{equation}
The drop in $\chi_l$ comes about because
the width of the distribution of ${\rm Re}(\rho_{\phi})$ becomes much smaller
below the transition.
At low temperatures structural arrest hinders the exploration of phase space
and reduces the fluctuations in Re$(\rho_{\phi})$.

The other way to calculate the generalized linear compressibility
is with global averaging in which we string together a series of separate runs,
treat it as one giant run, and then do the averaging required to calculate
the generalized linear compressibility $\chi_{l}^{\rm global}$. 
\begin{equation}
\chi_{l}^{\rm global}=\frac{6}{N}
\left\{\left[\langle\rho_{\phi}^2\rangle\right]_{\rm run}-
\left(\left[\langle\rho_{\phi}\rangle\right]_{\rm run}\right)^2\right\}
\label{eq:globalavg}
\end{equation}
Careful inspection of eqs. (\ref{eq:runbyrunavg}) and (\ref{eq:globalavg})
reveals that the difference is in whether we square the thermal average
$\langle\rho_{\phi}\rangle$ and then average over runs to obtain
$\left[\langle\rho_{\phi}\rangle^2\right]_{\rm run}$ in $\chi_l$, or
average $\langle\rho_{\phi}\rangle$ over runs and then square it to obtain
$\left(\left[\langle\rho_{\phi}\rangle\right]_{\rm run}\right)^2$ in
$\chi_{l}^{\rm global}$.
If we now return to Figure \ref{fig:Sk} and take
the difference of $[S(k)]_{\rm run}$
and $\left([{\rm Re}<\rho_{k}>]_{\rm run}\right)^2$, 
we obtain the global average:
\begin{equation}
\chi_{l}^{\rm global}=3[S(k)]_{\rm run}-
\frac{6}{N}\left([{\rm Re}<\rho_{k}>]_{\rm run}\right)^2 
\end{equation}
The result of both types of averaging is shown in Figures \ref{fig:Sk} 
and \ref{fig:linGlobal}. Notice that $\chi_{l}^{\rm global}$ does
not exhibit a drop with decreasing temperature while $\chi_{l}$ does.
To understand why there is no drop in $\chi_{l}^{\rm global}$, note
that by combining several different runs, very different configurations
are sampled which produces much larger fluctuations in the generalized 
center of mass at low temperatures compared to $\chi_{l}$. As a result
$\chi_{l}^{\rm global}$, which is a measure of the size of
these fluctuations, does not have an abrupt drop.

\begin{Fig}{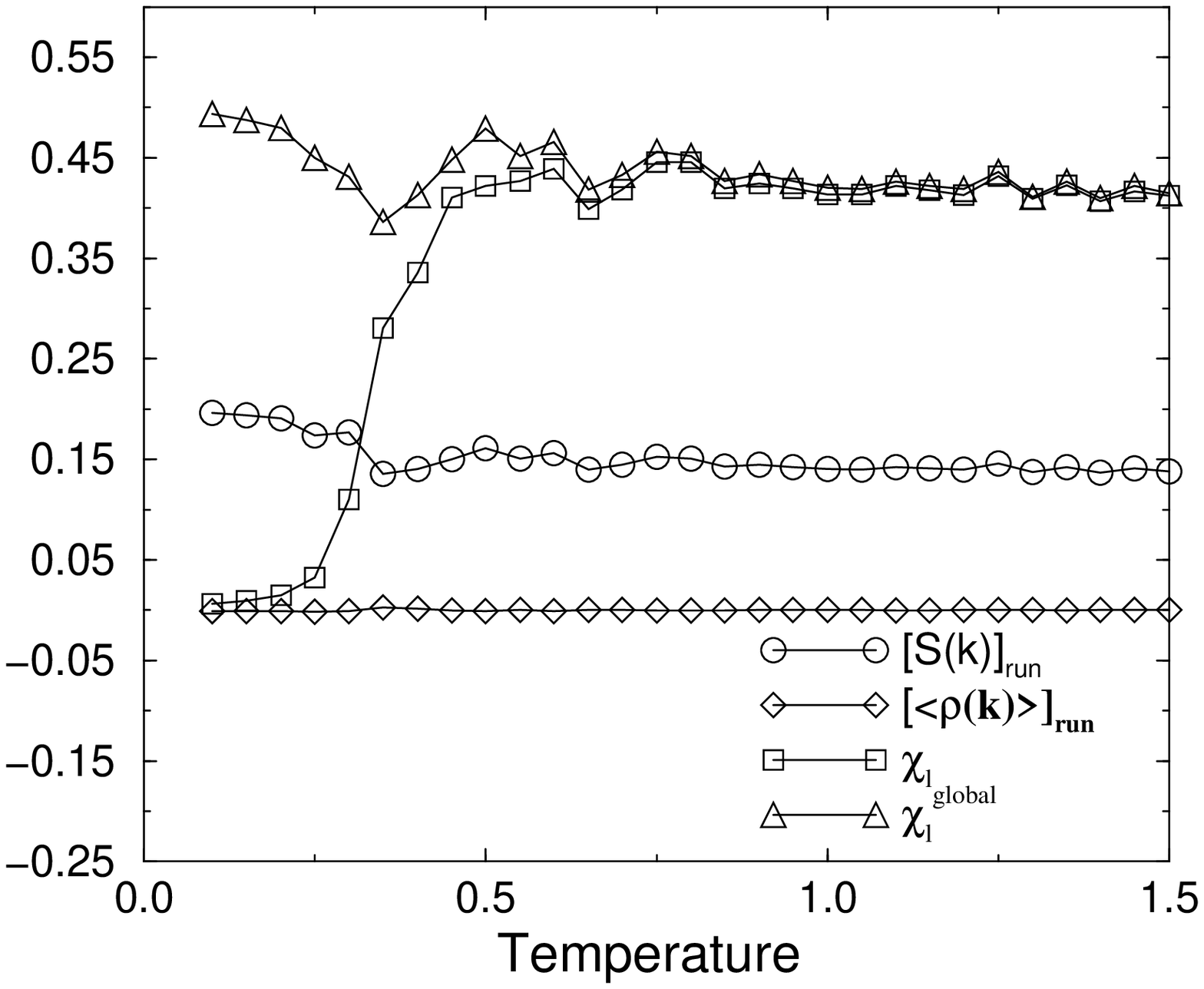}
\caption{Linear generalized compressibility $\chi_l$, static structure factor 
$[S(k)]_{\rm run}$, $[{\rm Re}<\rho_{k}>]_{\rm run}$,
linear generalized compressibility $\chi_l$ averaged run by run,
and the globally averaged linear generalized compressibility 
$\chi_l^{\rm global}$ versus temperature for a
binary mixture of 512 particles. The measurement time was $10^6$ md steps 
for each temperature. The data was averaged over 10 runs and over 
$\vec{k}=(2\pi/L,0,0)$, $\vec{k}=(0,2\pi/L,0)$, and $\vec{k}=(0,0,2\pi/L)$.
$\rho_{o}=0.6$ and $\sigma_{B}/\sigma_{A}=1.4$.}
\label{fig:Sk}
\end{Fig}

\begin{Fig}{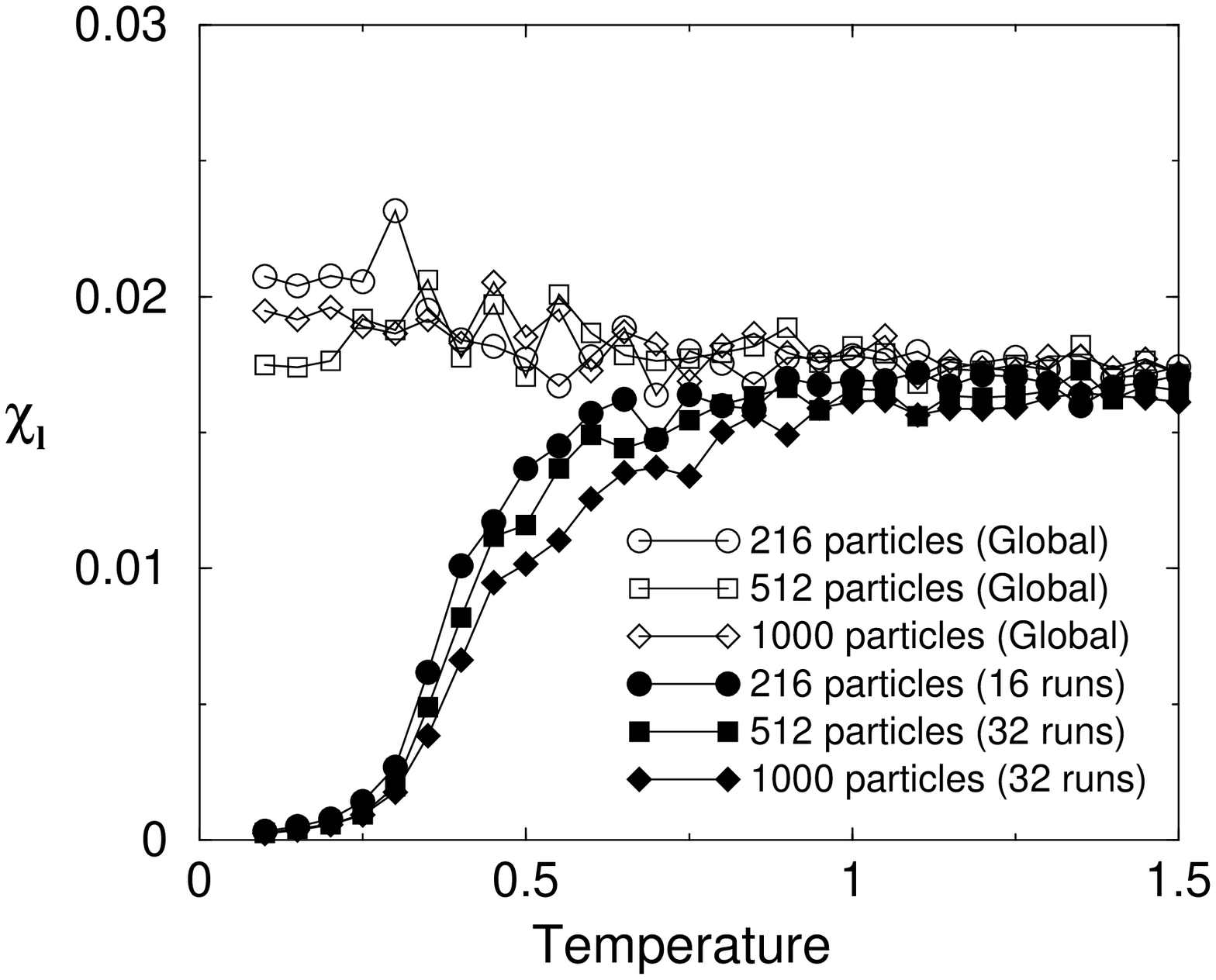}
\caption{Linear generalized compressibility as a function of
temperature for binary mixtures of 216, 512 and 1000 particles. 
The filled symbols correspond to calculating $\chi_l$
for each run with a measuring time of $2\times 10^5$ md steps and then
averaging over the number of runs indicated in the legend, while
the open symbols correspond to global averages (stringing all these
runs together to get one big ``run'' for a given system size). So the
global average for 216 particles uses $3.2 \times 10^6$ md steps while for
512 and 1000 particles, $6.4 \times 10^6$ md steps were used. 
$\chi_{l}$ is calculated using eqs. (\protect\ref{eq:susceptibilities})
and (\protect\ref{eq:abs}). $\rho_o = 0.6$ and $\sigma_B/\sigma_A=1.4$.}
\label{fig:linGlobal}
\end{Fig}

\section{Summary}
To summarize, we have introduced a new quantity called the generalized
compressibility which depends solely on the positions of the particles 
and not on their histories. The generalized compressibility can easily be 
calculated in the canonical (e.g., $NVT$) and grand canonical ensembles. 
In particular it is well defined in a system which has particle
number and volume fixed. In addition it does not suffer from the
finite size effects often encountered in calculating the ordinary
compressibility. The linear generalized compressibility  
drops abruptly at the observed glass transition due to the kinetic arrest
of motion. This makes it an good quantity to calculate
or measure in order to find the observed glass transition as
a function of density or temperature.
The generalized compressibility can be experimentally
measured in several ways. It can be directly measured in colloidal
experiments which monitor the positions of the particles. 
Measurements of the width of the distribution of
$\rho_{\vec{k}}$, the spatial Fourier transform of the density, would also
yield the linear generalized compressibility.
 
We thank Sharon Glotzer, Walter Kob, Andrea Liu and Francesco Sciortino 
for helpful discussions.
This work was supported in part by DOE grant DE-FG03-00ER45843 as well as by
CULAR funds provided by the University of California for the conduct of
discretionary research by Los Alamos National Laboratory.

\section{Appendix: Parallel Tempering}
In calculating the intermediate scattering function at a given
temperature, we initialized the run using a configuration at that
temperature generated by parallel tempering. In this appendix
we briefly describe the parallel tempering method.

We implement parallel tempering (PT) 
\cite{Hukushima96,Hukushima98,Kob01,Michele01} by choosing the
temperatures at which we wish to have measurements made. We then run
molecular dynamics simulations in parallel at these temperatures using
a temperature constraint algorithm \cite{Rapaport95} to keep the 
temperature of each simulation constant. 
At 100 time step intervals we attempt to switch the configurations
of two neighboring temperatures using a Metropolis test which ensures
that the energies of the configurations sampled at any given 
temperature have a Boltzmann distribution. Let $\beta_1$ and $\beta_2$
be two neigboring inverse temperatures, and let $U_1$ and $U_2$ be
the corresponding potential energies of the configurations at these
temperatures at a time step just before the possible swap. If
$\Delta=(\beta_1 - \beta_2)(U_2 - U_1)$, then the switch is accepted
with probability unity if $\Delta\leq 0$ and with probability 
$\exp(-\Delta)$ if $\Delta >0$. The temperatures are chosen so that
the acceptance ratio is between
30\% and 75\%. At the temperatures in the vicinity of the mode coupling
$T_C$, the acceptance ratio was typically above 75\% for
L=6 and above 60\% for L=8. After a swap is accepted, the velocities
of the particles in each configuration are rescaled to suit their 
new temperature. Each configuration is then evolved using molecular
dynamics for another 100 time steps. Switching configurations 
allows a given
simulation to do a random walk in temperature space in which it
visits both low temperatures and high temperatures. This helps to
prevent a simulation from becoming trapped in a valley of the energy
landscape at low temperatures. Typically we equilibrate for 
2 million time steps and then do measurements for an additional
4 million time steps. 
 
$^{\dagger}$Present address: Internap, Seattle, WA 98101.\\

\end{document}